\documentclass[twocolumn,showpacs,preprintnumbers,superscriptaddress,amsmath,amssymb,pre]{revtex4}

\usepackage{graphicx}
\usepackage{dcolumn}
\usepackage{bm}

\begin{document}

\preprint{Submitted to {\em{Physical Review E}}}

\title{Statistical significance of rich-club phenomena in complex networks}

\author{Zhi-Qiang Jiang}
 \affiliation{School of Business, East China University of Science and Technology, Shanghai 200237, China}
 \affiliation{School of Science, East China University of Science and Technology, Shanghai 200237, China}

\author{Wei-Xing Zhou}
 \email{wxzhou@ecust.edu.cn}
 \affiliation{School of Business, East China University of Science and Technology, Shanghai 200237, China}
 \affiliation{School of Science, East China University of Science and Technology, Shanghai 200237, China}
 \affiliation{Research Center of Systems Engineering, East China University of Science and Technology, Shanghai 200237, China}

\date{\today}

\begin{abstract}
We propose that the rich-club phenomena in complex networks should
be defined in the spirit of bootstrapping, in which a null model is
adopted to assess the statistical significance of the rich-club
detected. Our method can be served as a definition of rich-club
phenomenon and is applied to analyzing three real networks and three
model networks. The results improve significantly compared with
previously reported results. We report a dilemma with an exceptional
example, showing that there does not exist an omnipotent definition
for the rich-club phenomenon.
\end{abstract}


\maketitle

Almost all social and natural systems are composed of a huge number
of interacting components. Many self-organized features that are
absent at the microscopic level emerge in complex systems due to the
dynamics. The topological properties of the underlying network of
the interacting constituents have great impact on the dynamics of
the system evolving on it
\cite{Albert-Barabasi-2002-RMP,Newman-2003-SIAMR,Dorogovtsev-Mendes-2003,Boccaletti-Latora-Moreno-Chavez-Hwang-2006-PR}.
Most complex networks exhibit small-world properties
\cite{Watts-Strogatz-1998-Nature} and are scale free in the sense
that the distribution of degrees has power-law tails
\cite{Barabasi-Albert-1999-Science}. In addition, many real networks
have modular structures or communities expressing their underlying
functional modules \cite{Newman-2004-EPJB} and exhibit self-similar
and scale invariant nature in the topology
\cite{Kim-2004-PRL,Song-Havlin-Makse-2005-Nature,Strogatz-2005-Nature,Song-Havlin-Makse-2006-NP,Goh-Salvi-Kahng-Kim-2006-PRL,Kim-Goh-Salvi-Oh-Kahng-Kim-2006-XXX,Zhou-Jiang-Sornette-2007-PA}.
The modular and hierarchical structure of social networks may partly
account for the log-periodic power-law patterns presented
extensively in financial bubbles and antibubbles
\cite{Sornette-2003-PR,Zhou-Sornette-Hill-Dunbar-2005-PRSB}. A
closely relevant feature is recently reported in some complex
networks, termed the rich-club phenomenon, which however lacks a
consensus on its definition
\cite{Zhou-Mondragon-2004-PRE,Zhou-Mondragon-2004-IEEE,Zhou-Mondragon-2005-LNCS,Zhou-Zhang-Zhang-2005-XXX,Colizza-Flammini-Serrano-Vespignani-2006-NP,Amaral-Guimera-2006-NP}.

The rich-club phenomenon in complex networks digests the observation
that the nodes with high degree (called rich nodes) are inclined to
intensely connect with each other. The average hop distance of the
tight group is between one and two \cite{Zhou-Mondragon-2004-IEEE}.
Intuitively, rich nodes are much more likely to organize into tight
and highly-interconnected groups (clubs) than low-degree nodes.
Therefore, it is rational to accept that there is a rich-club
phenomenon in the topology of internet
\cite{Zhou-Mondragon-2004-PRE,Zhou-Mondragon-2004-IEEE,Zhou-Mondragon-2005-LNCS,Zhou-Zhang-Zhang-2005-XXX}.
This rationale can be characterized quantitatively by the rich-club
coefficient $\phi$, which is expressed as follows
\cite{Zhou-Mondragon-2004-IEEE},
\begin{equation}
\phi(k) = \frac{2E_{>k}}{N_{>k}(N_{>k}-1)},
 \label{Eq:RCC}
\end{equation}
where $N_{>k}$ refers to the number of nodes with the degrees higher
than a given value $k$ and $E_{>k}$ stands for the number of edges
among the $N_{>k}$ nodes. The rich-club coefficient $\phi(k)$ is the
ratio of the real number to the maximally possible number of edges
linking the $N_{>k}$ nodes, which measures how well the rich nodes
`know' each other. For example, $\phi = 1$ means that the members
within the club form a full connected network. Indeed, $\phi(k)$ is
nothing but the well-known clustering coefficient of the rich club.

Zhou and Mondrag{\'o}n argue that an increasing function $\phi(k)$
with respect to $k$ provides evidence for the presence of rich-club
structure \cite{Zhou-Mondragon-2004-IEEE}. However, Colizza {\em{et
al.}} point out that a monotonic increase of $\phi(k)$ is not enough
to infer the presence of rich-club phenomenon since even random
networks generated from the ER model, the MR model and the BA model
have an increasing $\phi(k)$ with respect to $k$
\cite{Colizza-Flammini-Serrano-Vespignani-2006-NP}. Instead, the
rich-club coefficient $\phi(k)$ should be normalized by a reference
and the correct null model that can serve as a reference is the
maximally random networks with the same sequence of degrees as the
network under investigation
\cite{Colizza-Flammini-Serrano-Vespignani-2006-NP,Amaral-Guimera-2006-NP}.
The maximally random networks can be generated with the
chain-switching method
\cite{Maslov-Sneppen-2002-Science,Milo-Kashtan-Itzkovitz-Newman-Alon-2004-XXX}.
The normalized rich-club coefficient is defined by
\begin{equation}
\rho(k)= \phi(k)/\phi_{\rm{ran}}(k)~,\label{Eq:rho}
\end{equation}
where $\phi_{\rm{ran}}(k)$ is the average rich-club coefficient of
the maximally random networks
\cite{Colizza-Flammini-Serrano-Vespignani-2006-NP}. The actual
presence of rich-club phenomenon in a network is confirmed if
$\rho(k)>1$
\cite{Colizza-Flammini-Serrano-Vespignani-2006-NP,Amaral-Guimera-2006-NP}.
In this framework, there is no rich-club ordering in the network of
Internet.

\begin{figure*}[htb]
\centering
\begin{minipage}[t]{0.32\textwidth}
\includegraphics[width=5.4cm]{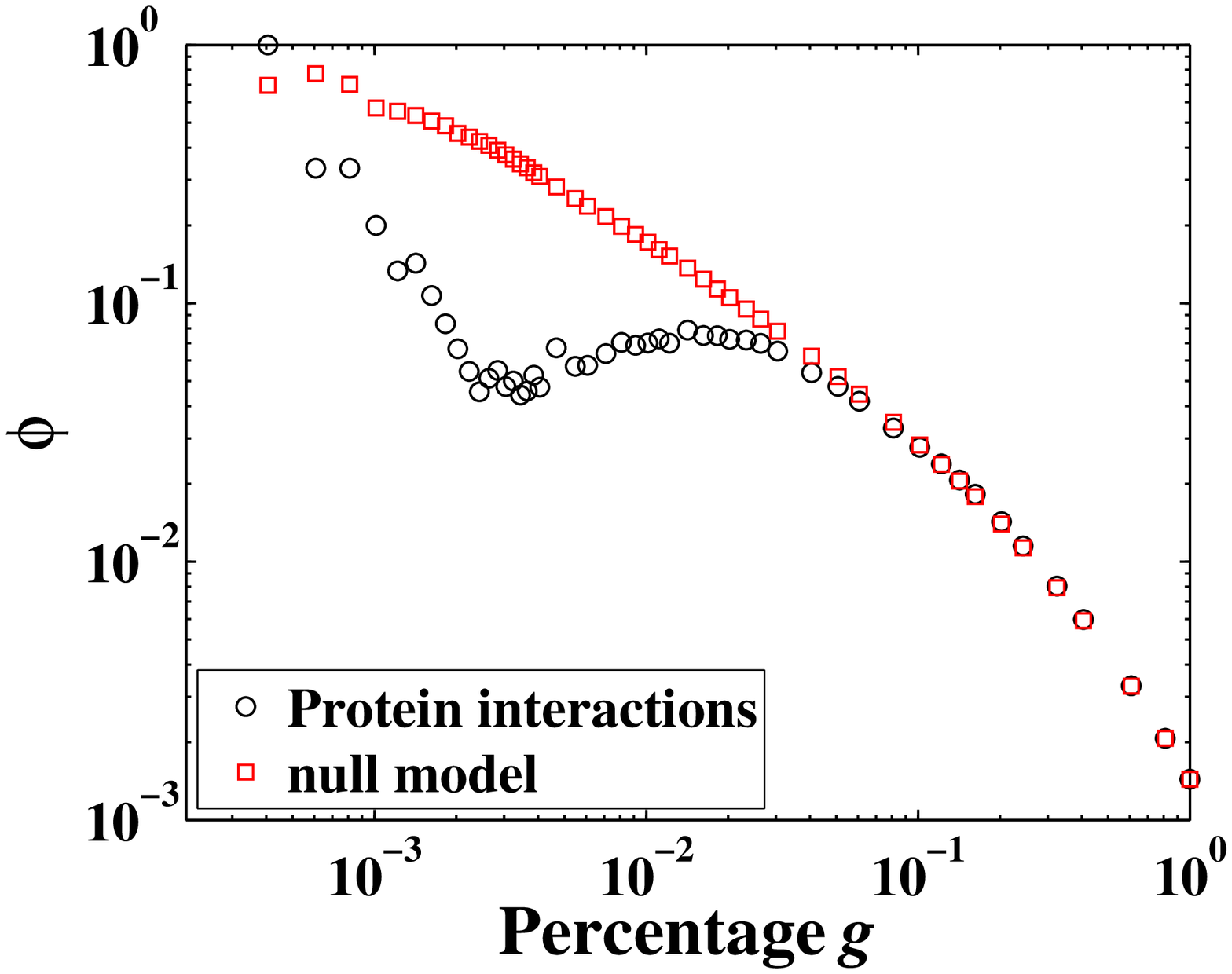}
\end{minipage}
\begin{minipage}[t]{0.32\textwidth}
\includegraphics[width=5.4cm]{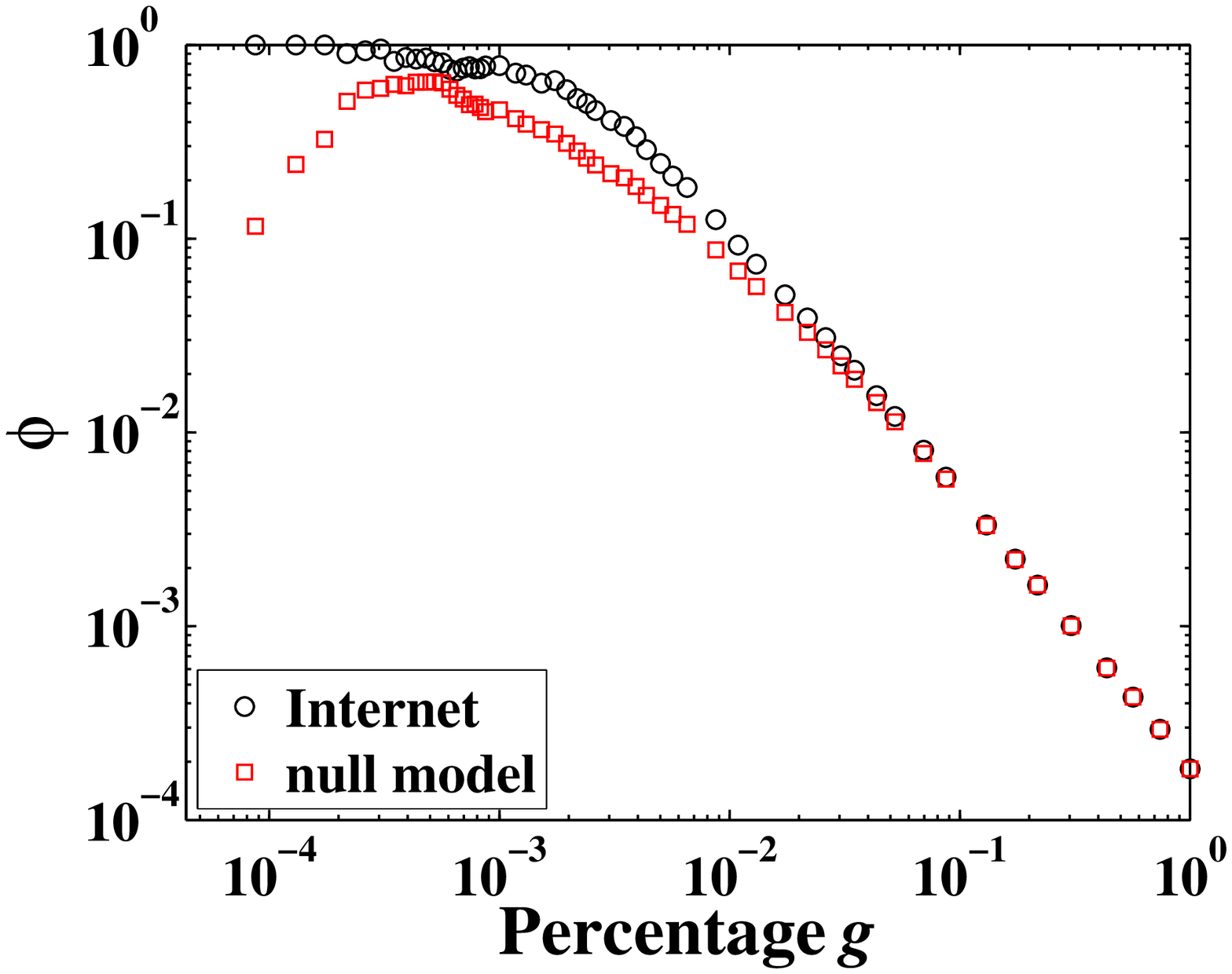}
\end{minipage}
\begin{minipage}[t]{0.32\textwidth}
\includegraphics[width=5.4cm]{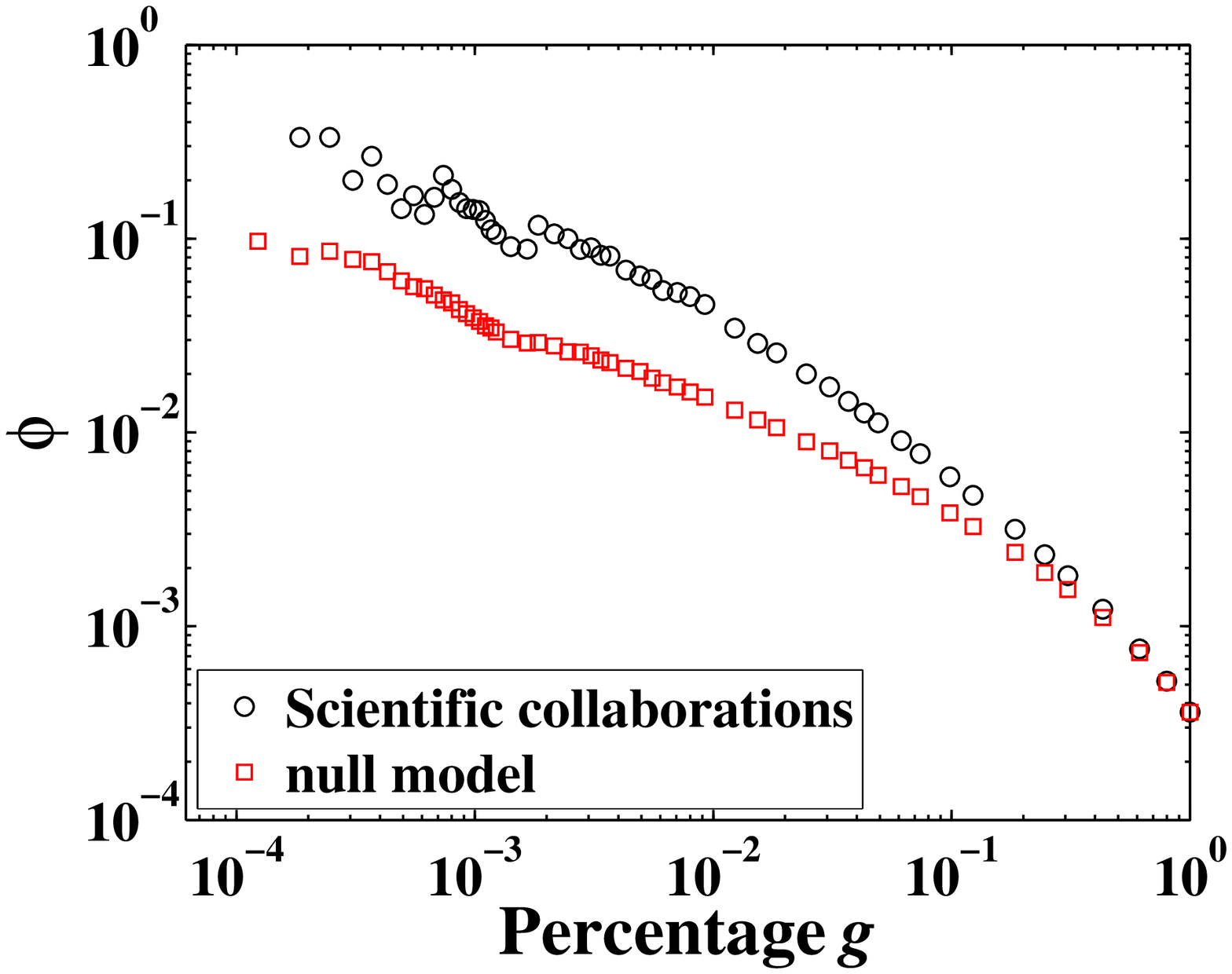}
\end{minipage}\\
\begin{minipage}[t]{0.32\textwidth}
\includegraphics[width=5.4cm]{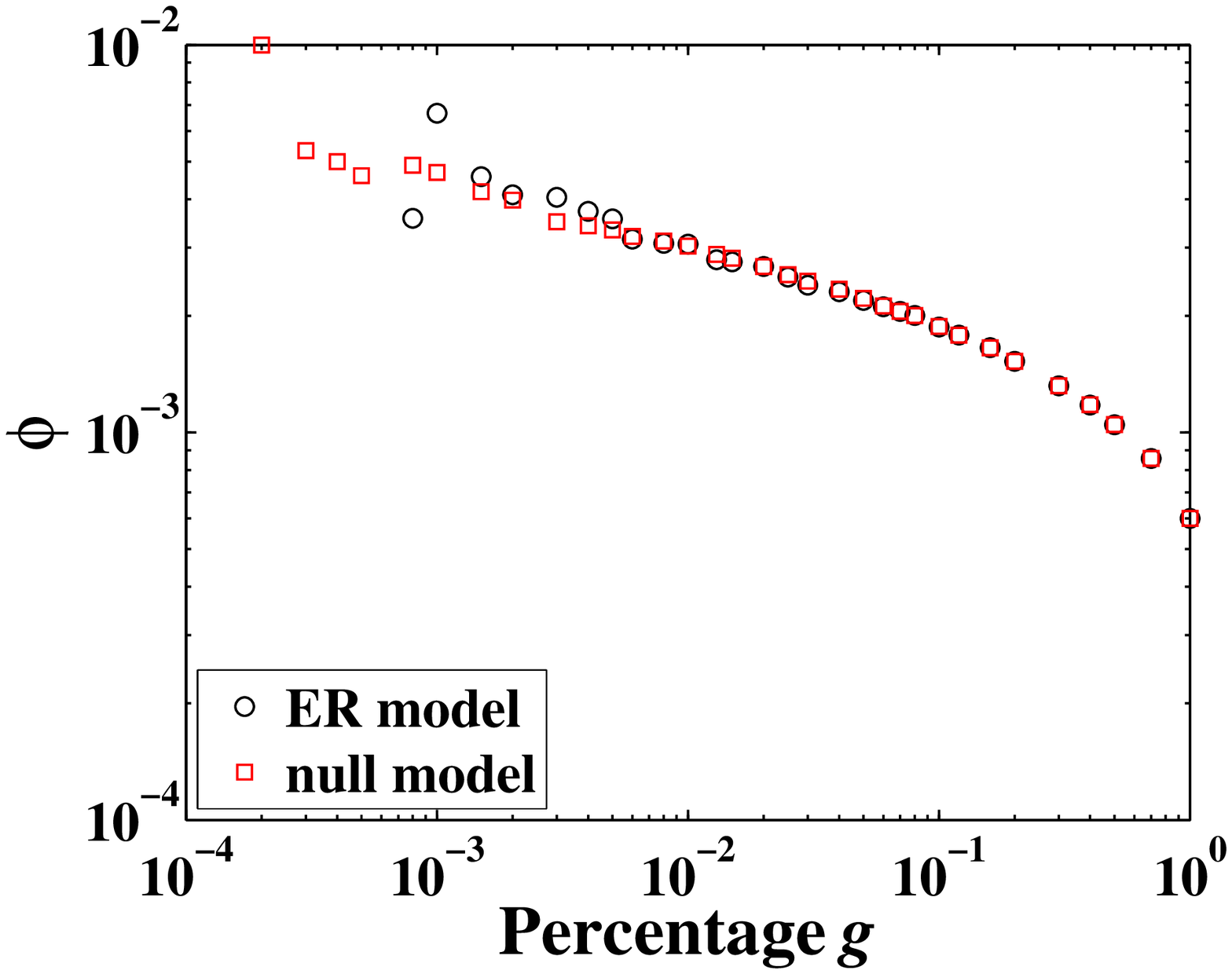}
\end{minipage}
\begin{minipage}[t]{0.32\textwidth}
\includegraphics[width=5.4cm]{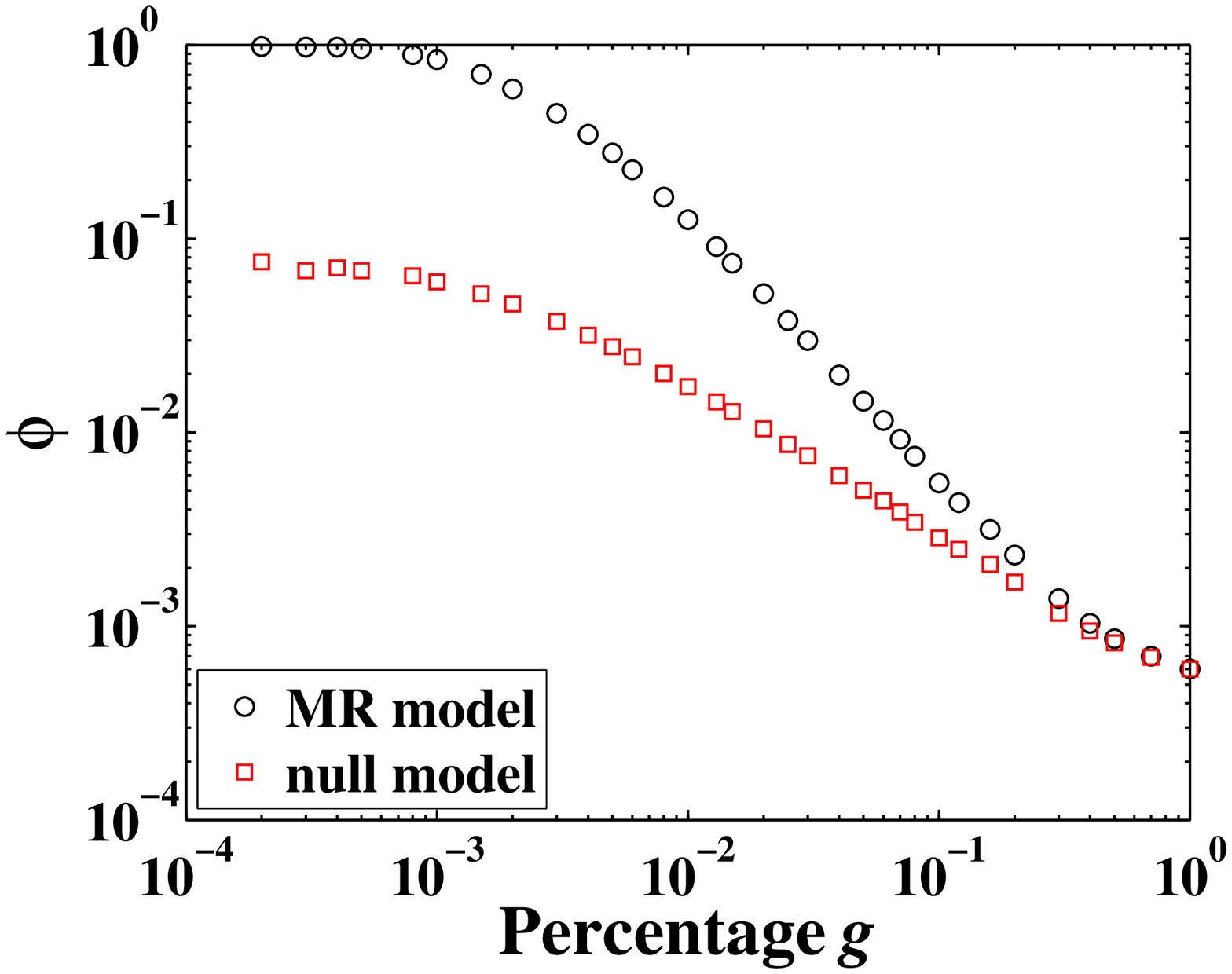}
\end{minipage}
\begin{minipage}[t]{0.32\textwidth}
\includegraphics[width=5.4cm]{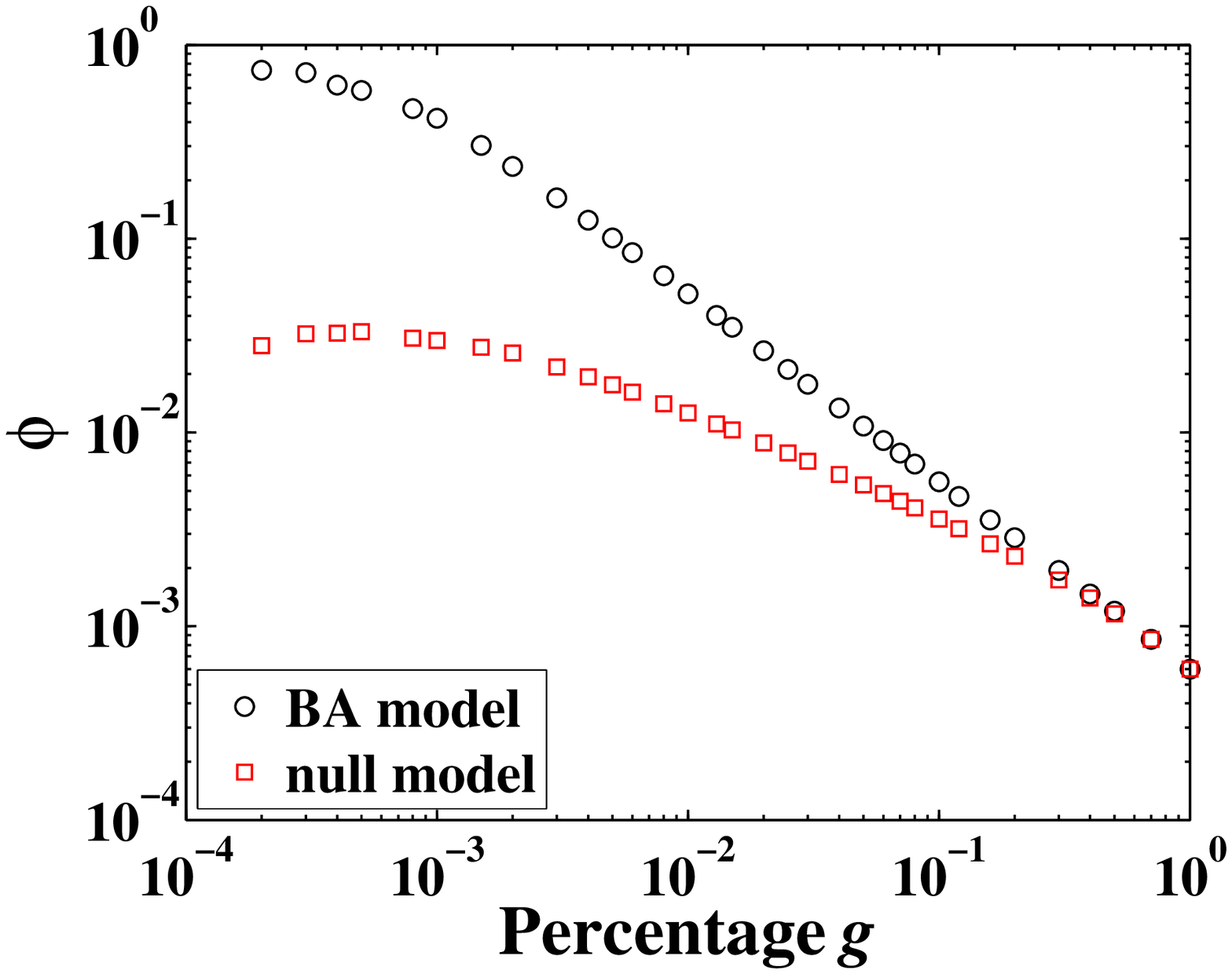}
\end{minipage}
\caption{(Color online) Rich-club coefficient $\phi$ as a function
of the percentage $g$ of nodes whose degree is larger than $k$ used
for detecting rich-club ordering. The black cycles are for the
networks under investigation and the red squares are for the null
models. For each simulated model network, the total number of nodes
is $10^4$ and its average degree is $\langle k \rangle = 6$. The
percentage of nodes with $k>1$ is $g=77.1\%$ for the protein
interaction network, $g=65.9\%$ for the Internet network, $g=86.6\%$
for the scientific collaboration network, $g=98.1\%$ for the ER
model, and there is no node with $k=1$ for the BA and MR networks.}
\label{Fig:phi}
\end{figure*}

We have repeated the analysis of Colizza {\em{et al}}
\cite{Colizza-Flammini-Serrano-Vespignani-2006-NP} for three model
networks, namely the Erd\"os-R\'enyi (ER) model
\cite{Erdos-Renyi-1959-PM}, the Molloy-Reed (MR) model
\cite{Molloy-Reed-1995-RSA}, and the Barab\'asi-Albert (BA) model
\cite{Barabasi-Albert-1999-Science}, and three real-world networks
being the protein interaction network
\cite{Maslov-Sneppen-2002-Science} of the yeast {\em{Saccharomyces
cerevisiae}}, the scientific collaboration network collected by
Newman \cite{Newman-2001-PNAS} and the Internet network at the
autonomous system level collected by the Oregon Route Views project
\cite{Faloutsos-Faloutsos-Faloutsos-1999-CCR,PastorSatorras-Vazquez-Vespignani-2001-PRL,Vazquez-PastorSatorras-Vespignani-2002-PRE}.
The rich-club coefficients $\phi$ of the six networks under
investigation are presented in Fig.~\ref{Fig:phi} with black circles
as a function of the percentage $g$ of the richest nodes included in
the rich club. The $\phi_{\rm{ran}}$ functions are also shown for
the corresponding maximally random networks. We note that when we
plot $\phi(k)$ versus $k$, our results for the investigated networks
are the same as shown in Fig. 1 obtained by Colizza {\em{et al}}
\cite{Colizza-Flammini-Serrano-Vespignani-2006-NP}.

%

\begin{figure*}[htb]
\centering
\begin{minipage}[t]{0.32\textwidth}
\includegraphics[width=5.4cm]{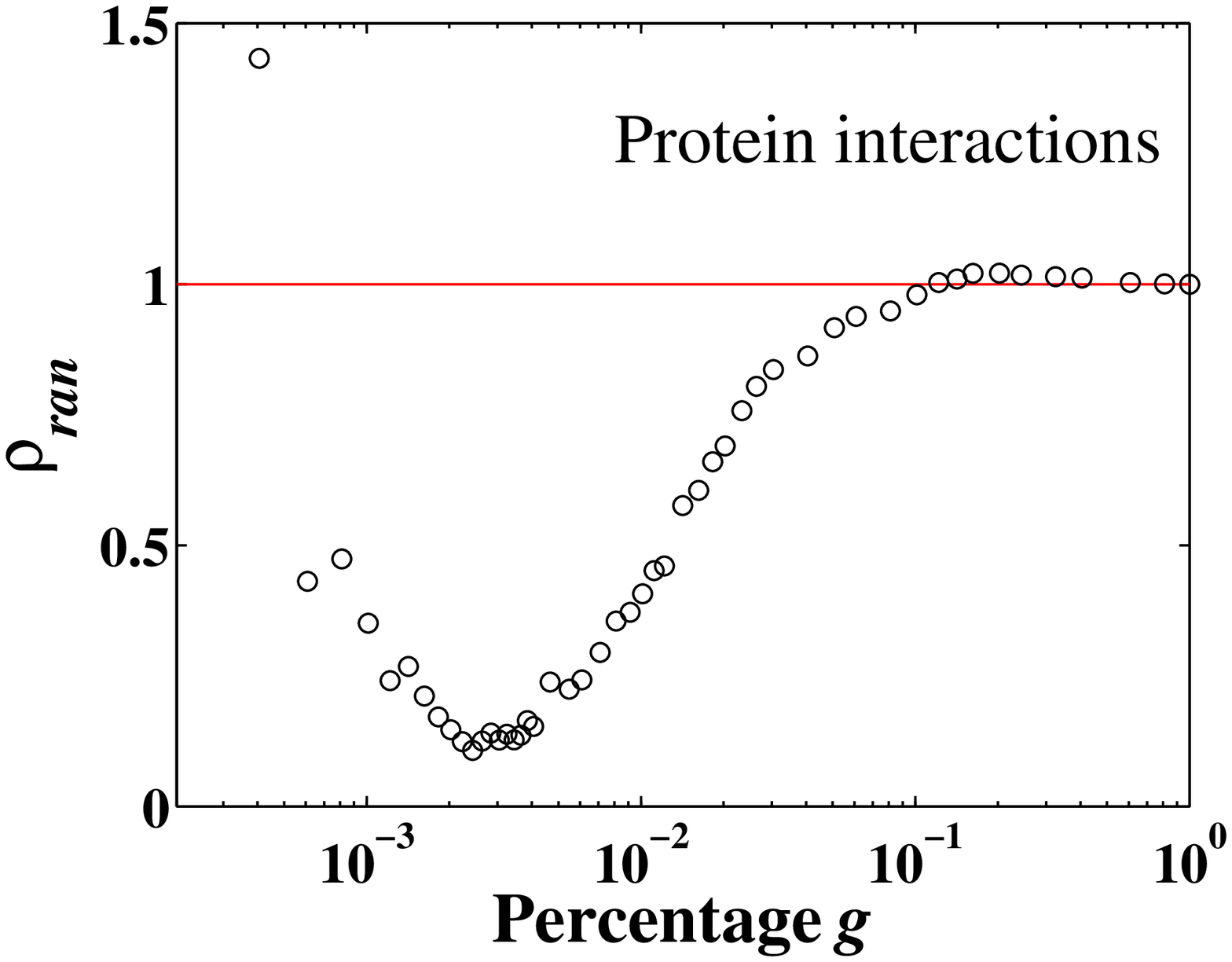}
\end{minipage}
\begin{minipage}[t]{0.32\textwidth}
\includegraphics[width=5.4cm]{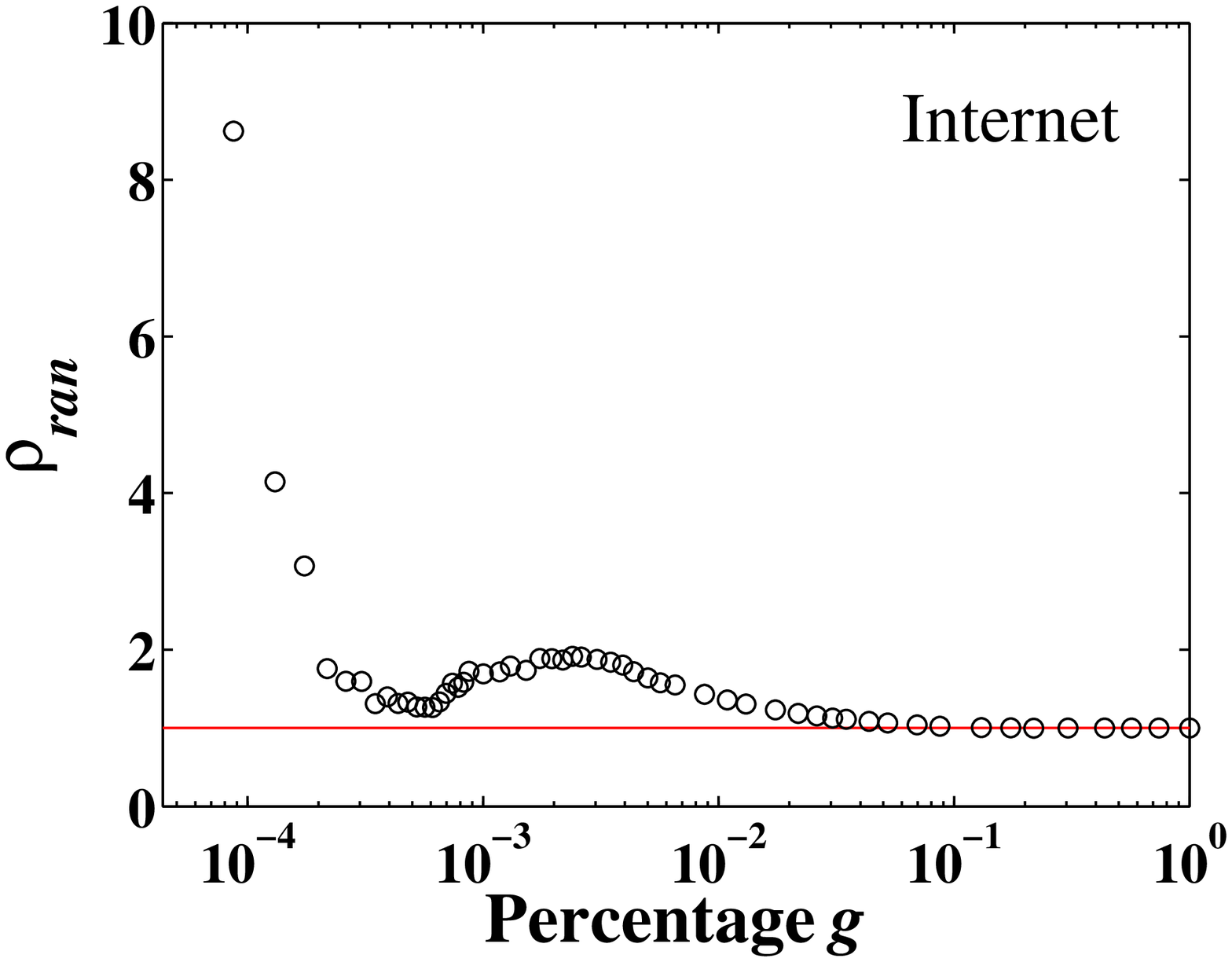}
\end{minipage}
\begin{minipage}[t]{0.32\textwidth}
\includegraphics[width=5.4cm]{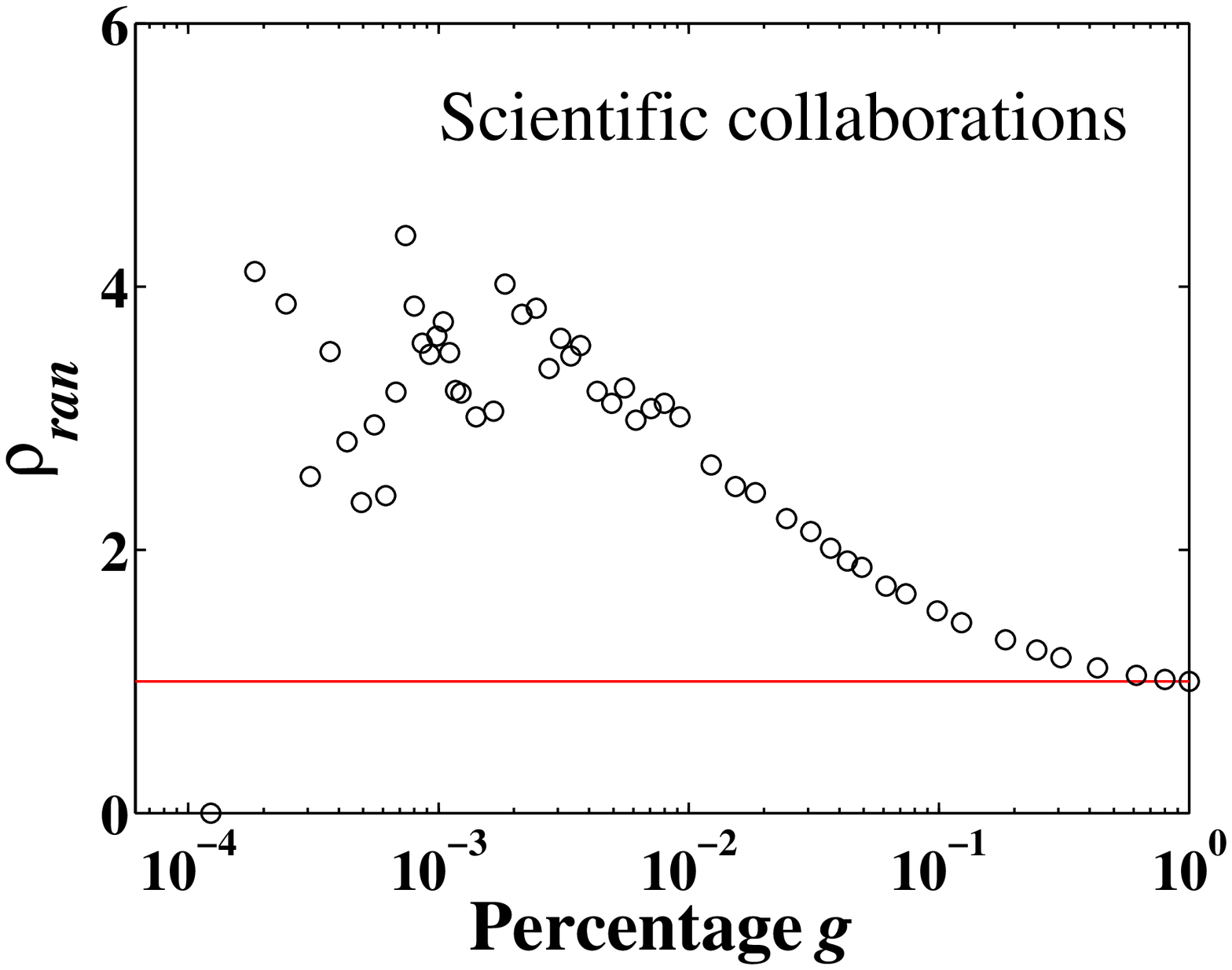}
\end{minipage}\\
\begin{minipage}[t]{0.32\textwidth}
\includegraphics[width=5.4cm]{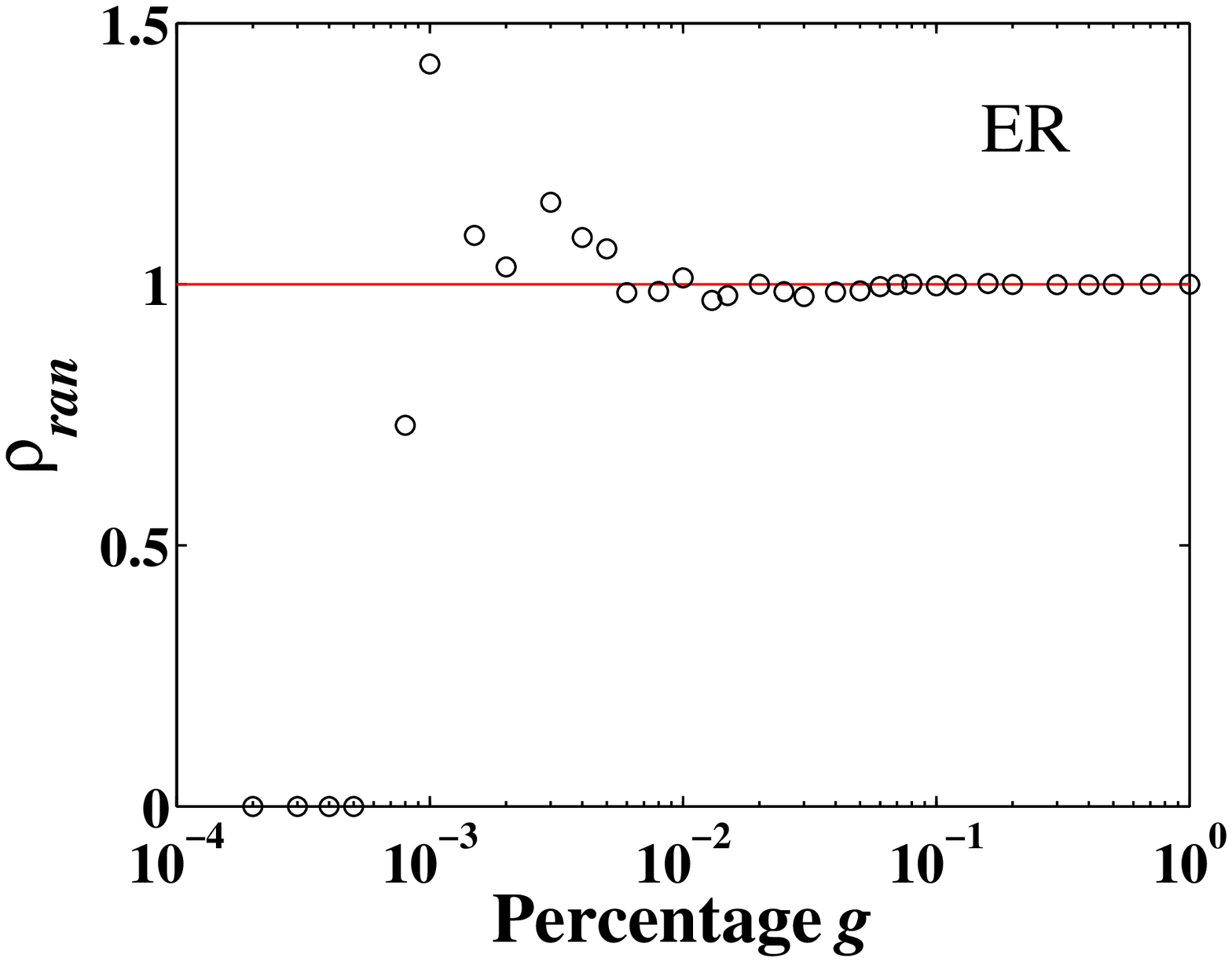}
\end{minipage}
\begin{minipage}[t]{0.32\textwidth}
\includegraphics[width=5.4cm]{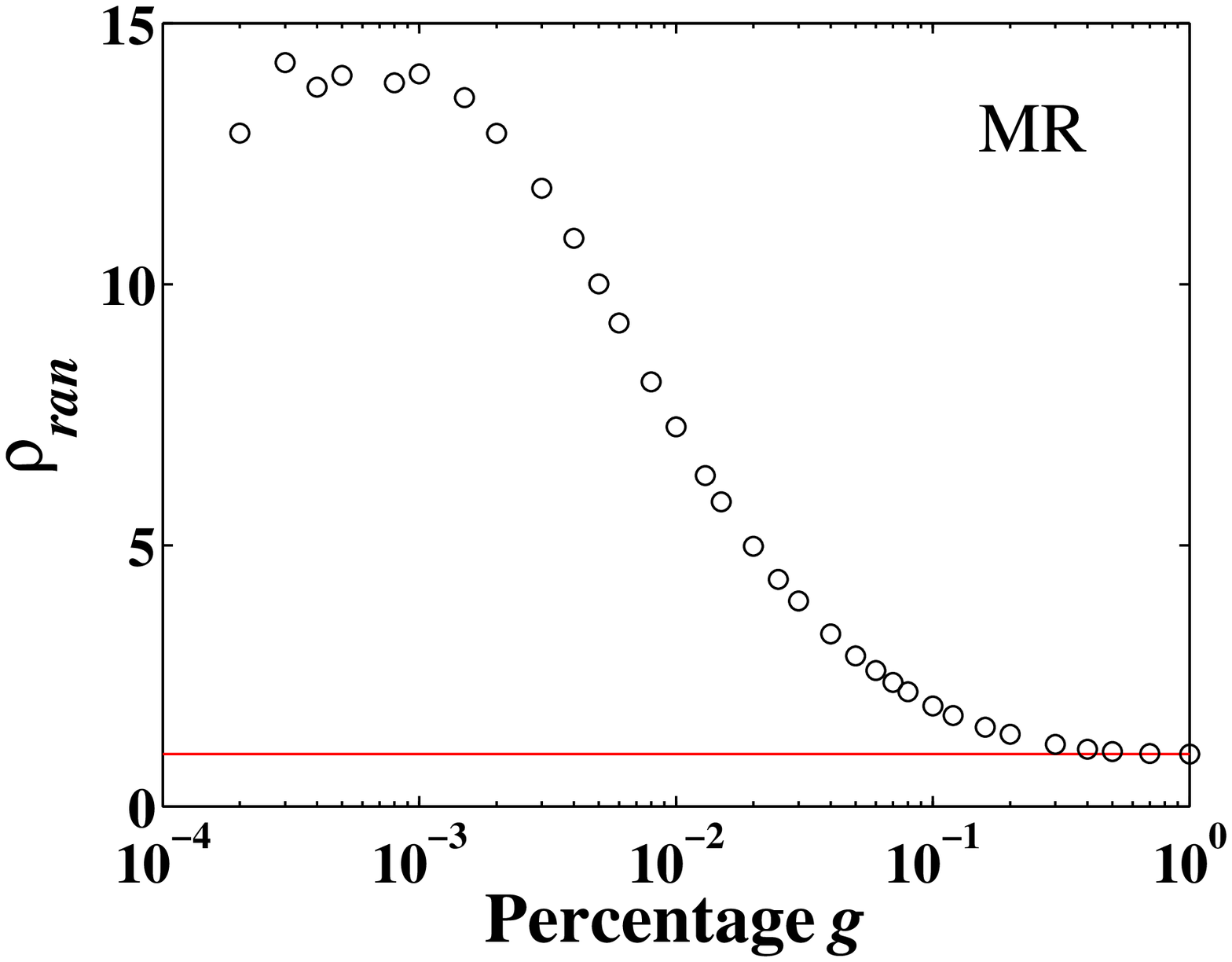}
\end{minipage}
\begin{minipage}[t]{0.32\textwidth}
\includegraphics[width=5.4cm]{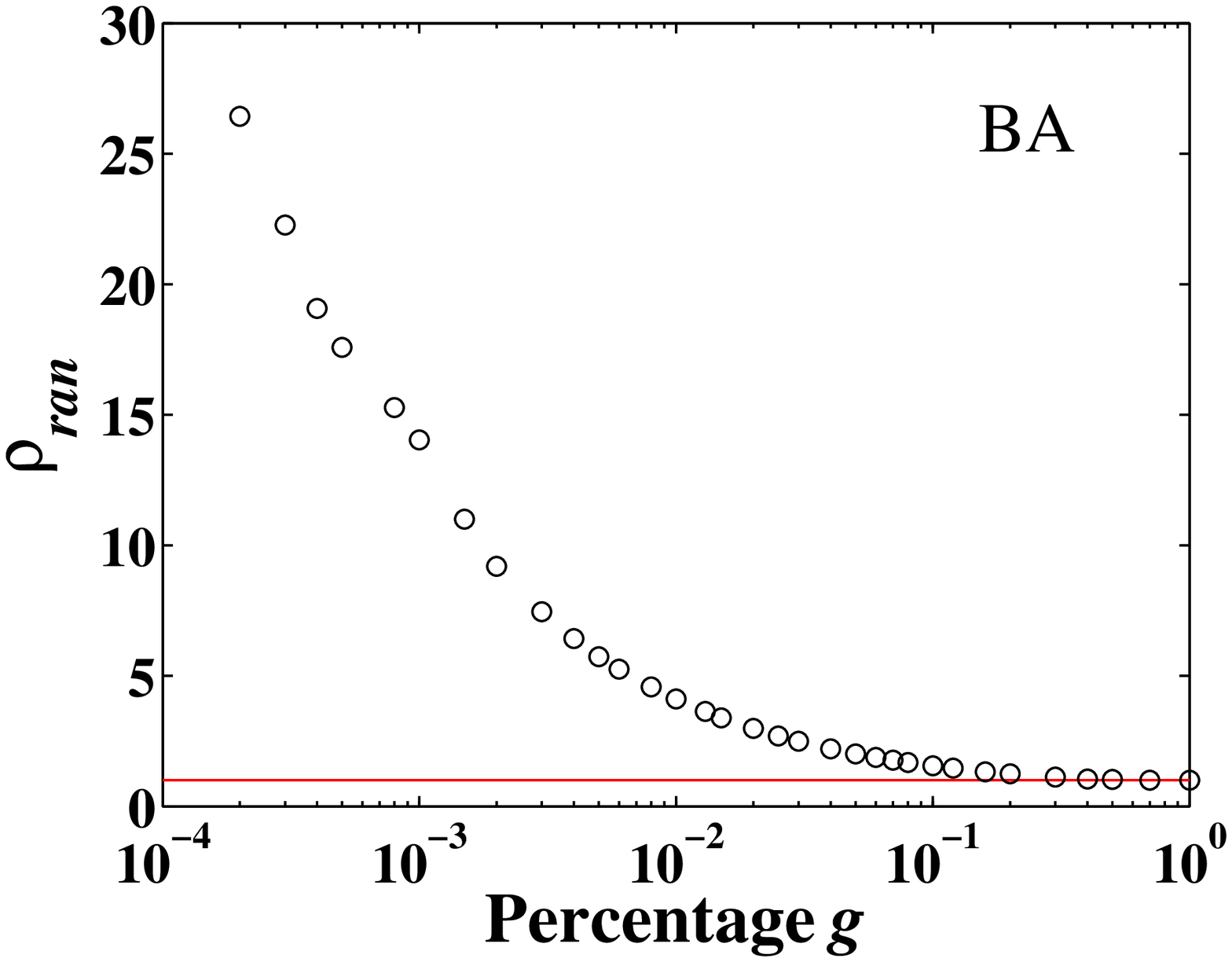}
\end{minipage}
\caption{Normalized rich-club coefficients of the investigated
networks. The ratio $\rho_{\rm{ran}} = \phi / \phi_{\rm{ran}}$ as a
function of the percentage $g$ and compared with the baseline value
equal to 1.} \label{Fig:rho}
\end{figure*}

Figure \ref{Fig:rho} shows the $\rho$ functions, which are not the
same as those in Fig.~2 presented by Colizza {\em{et al}}
\cite{Colizza-Flammini-Serrano-Vespignani-2006-NP}. Specifically, we
find that the normalized coefficients of the networks of protein
interactions, scientific collaborations, and the ER model are
qualitatively the same as those reported by Colizza {\em{et al}}
\cite{Colizza-Flammini-Serrano-Vespignani-2006-NP}, while the rest
three are not. We note that the AS-Internet data was created by Mark
Newman from data for July 22, 2006. Figure~\ref{Fig:rho} shows that
the normalized coefficient $\rho$ is not less than 1 for the
Internet, the MR model, and the BA model. For the Internet case, we
notice that its $\phi$ is close to 1 for the richest nodes.
Intuitively, the corresponding $\phi_{\rm{ran}}$ should be less than
1, which is observed in our analysis but not in that of Colizza
{\em{et al}} \cite{Colizza-Flammini-Serrano-Vespignani-2006-NP}.

%

The importance of null model has been emphasized in the assessment
of some properties claimed to be present in complex networks
\cite{Maslov-Sneppen-2002-Science,Amaral-Guimera-2006-NP,Guimera-SalesPardo-2006-MSB}.
Other than the simple normalization of the rich club coefficient, we
argue that the correct way to assess the presence of rich club
phenomenon is to perform a statistical test, which amounts to
determine the probability that the identified rich-club phenomenon
emerges by chance. The null hypothesis is the following:

{\em{$H_0$: $\rho(g)$ is not larger than 1.}}

\noindent The alternative hypothesis is that $\rho(g)>1$. We can
compute the $p$-value, which is the probability that the null
hypothesis is true. The smaller the $p$-value, the stronger the
evidence against the null hypothesis and favors the alternative
hypothesis that the presence of rich-club ordering is statistically
significant. The $p$-value is $100\%$ when $g=1$. By adopting the
conventional significance level of $\alpha=5\%$, the rich-club
phenomenon is statistically significant if $p<\alpha$.

\begin{figure*}[htb]
\centering
\begin{minipage}[t]{0.32\textwidth}
\includegraphics[width=5.4cm]{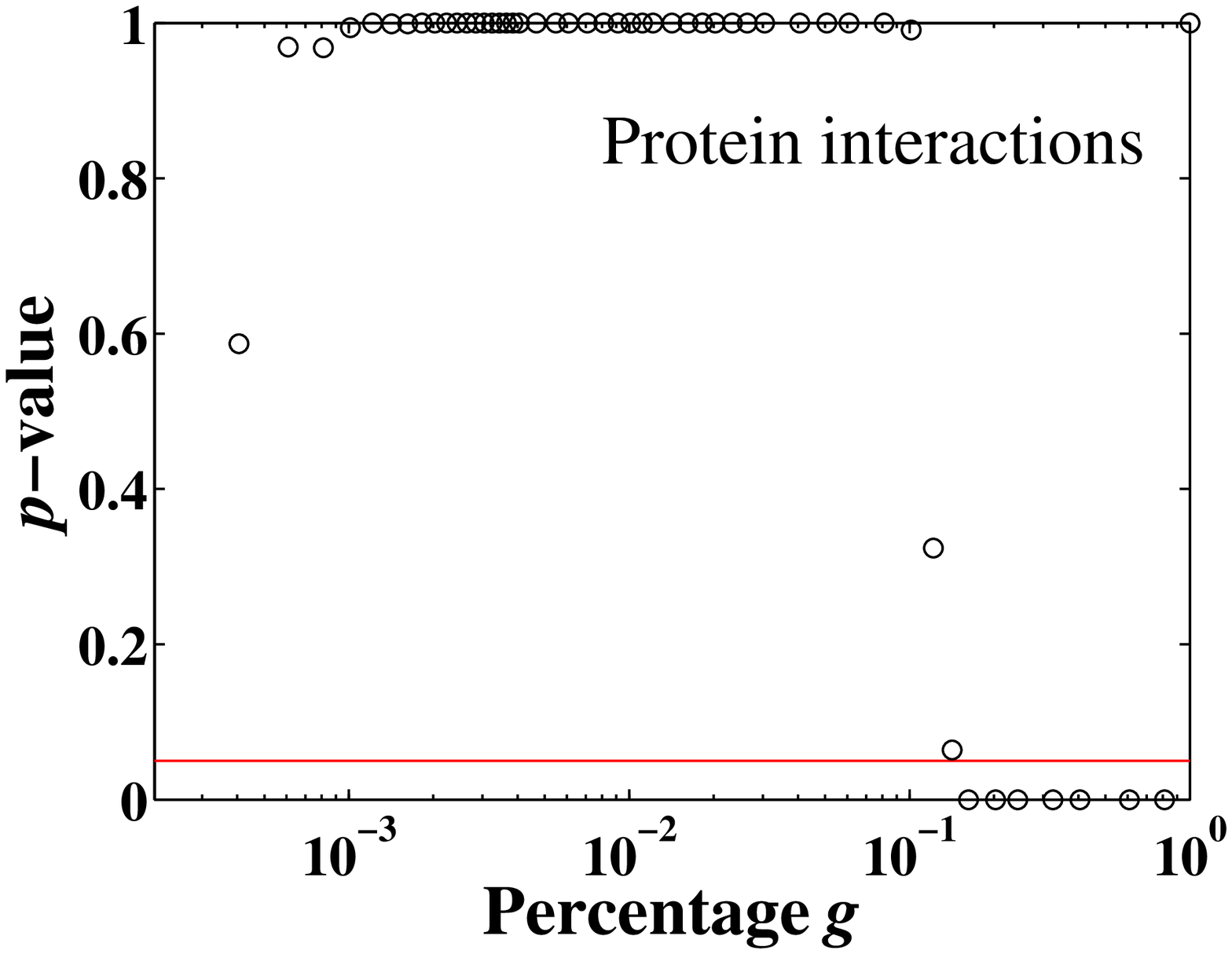}
\end{minipage}
\begin{minipage}[t]{0.32\textwidth}
\includegraphics[width=5.4cm]{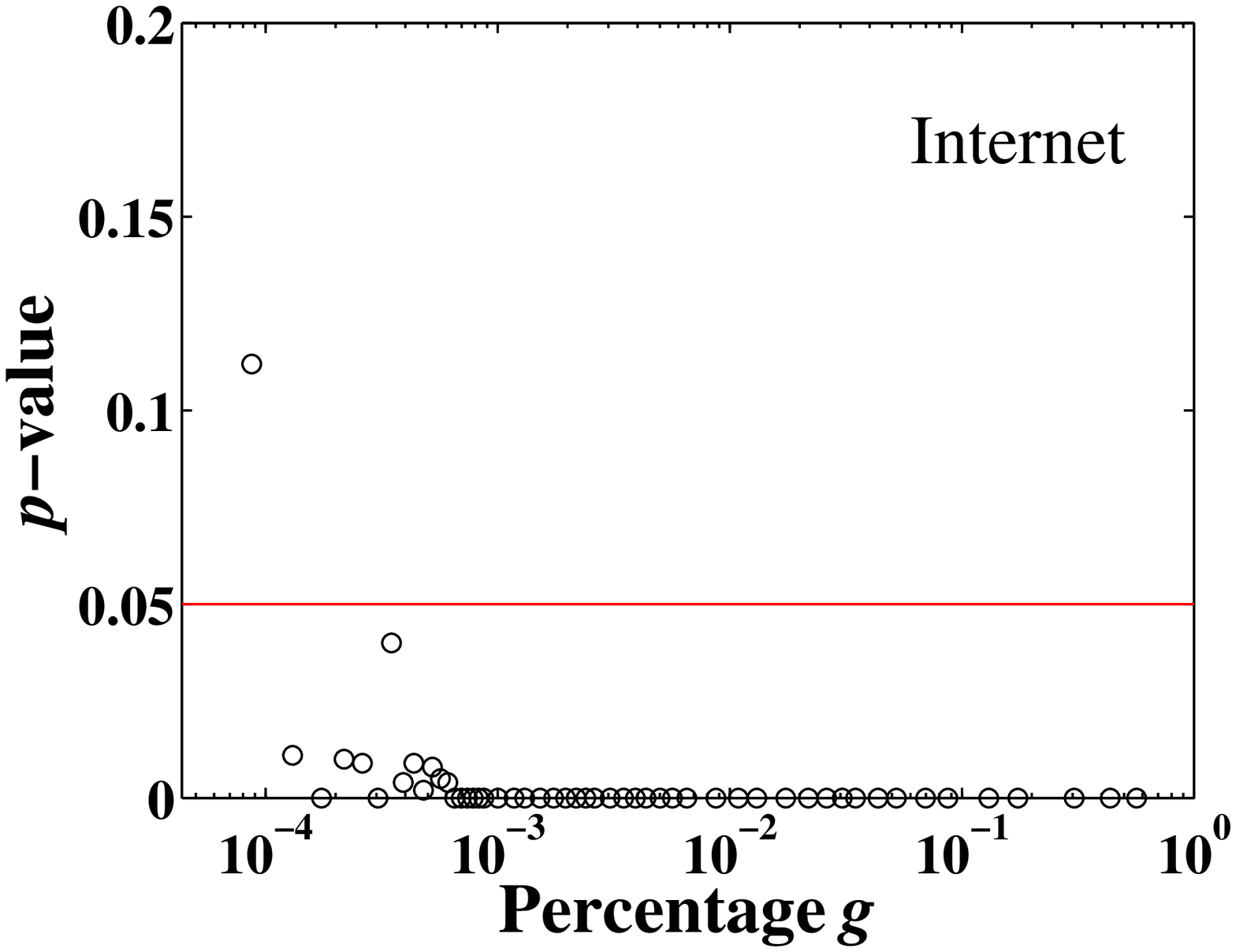}
\end{minipage}
\begin{minipage}[t]{0.32\textwidth}
\includegraphics[width=5.4cm]{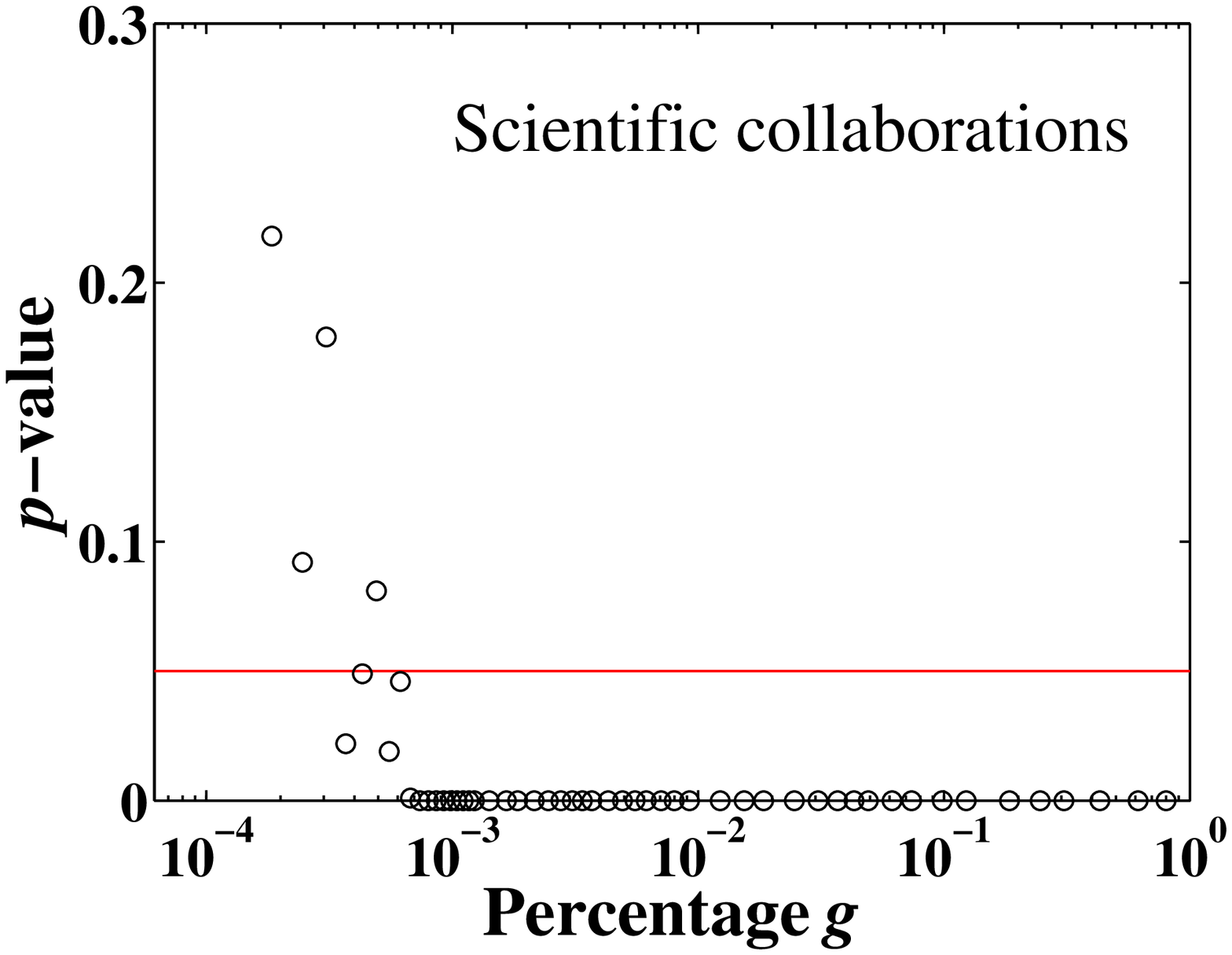}
\end{minipage}\\
\begin{minipage}[t]{0.32\textwidth}
\includegraphics[width=5.4cm]{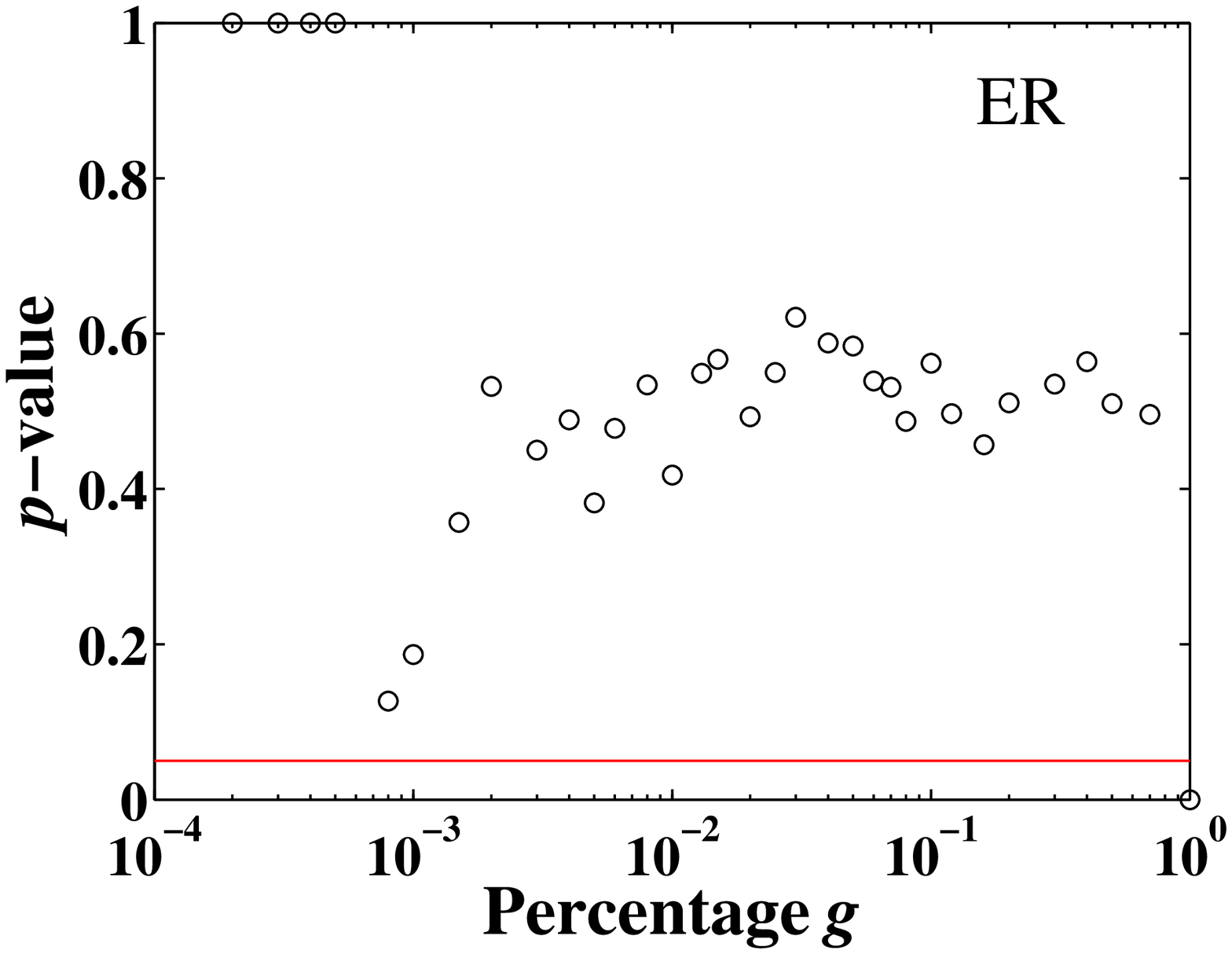}
\end{minipage}
\begin{minipage}[t]{0.32\textwidth}
\includegraphics[width=5.4cm]{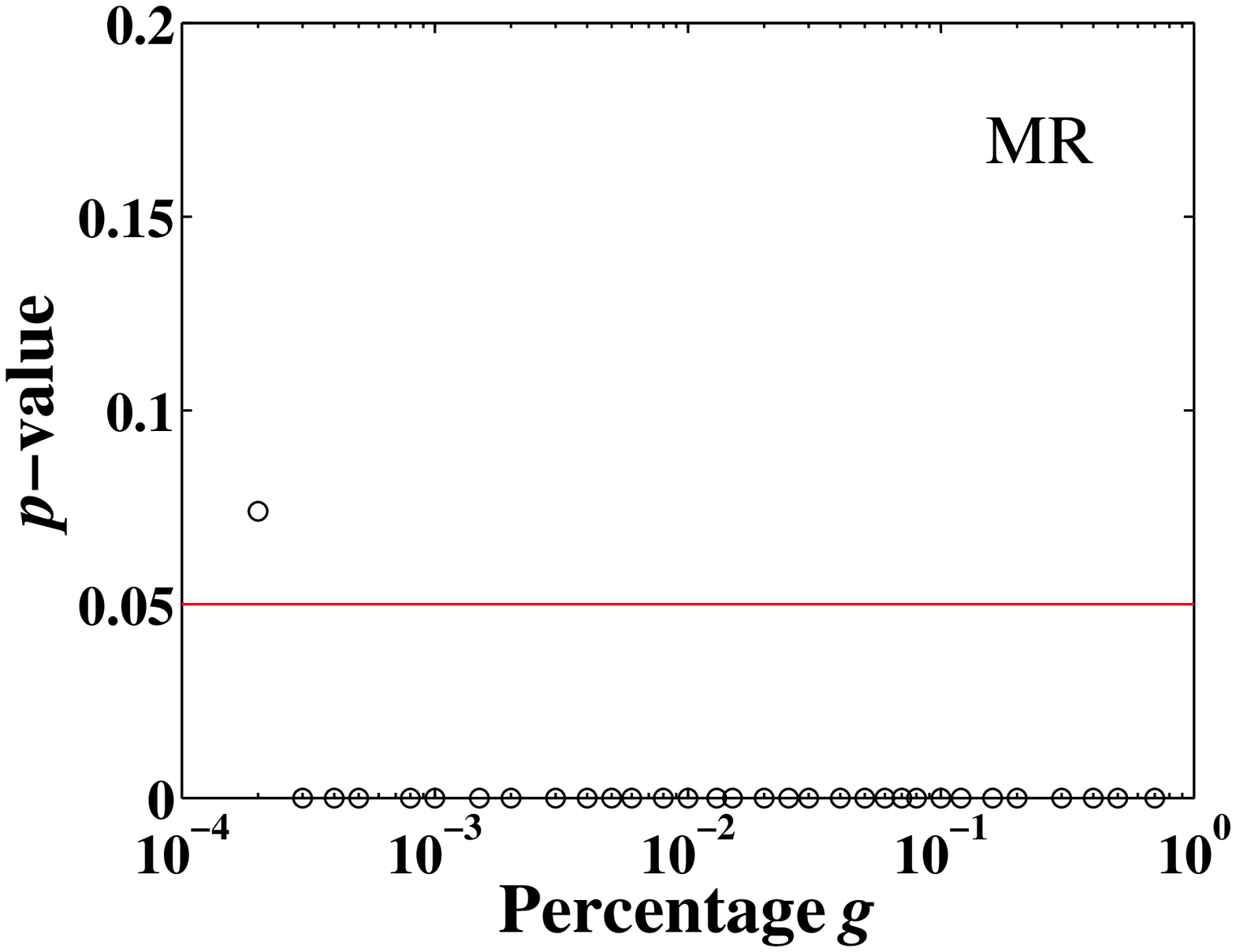}
\end{minipage}
\begin{minipage}[t]{0.32\textwidth}
\includegraphics[width=5.4cm]{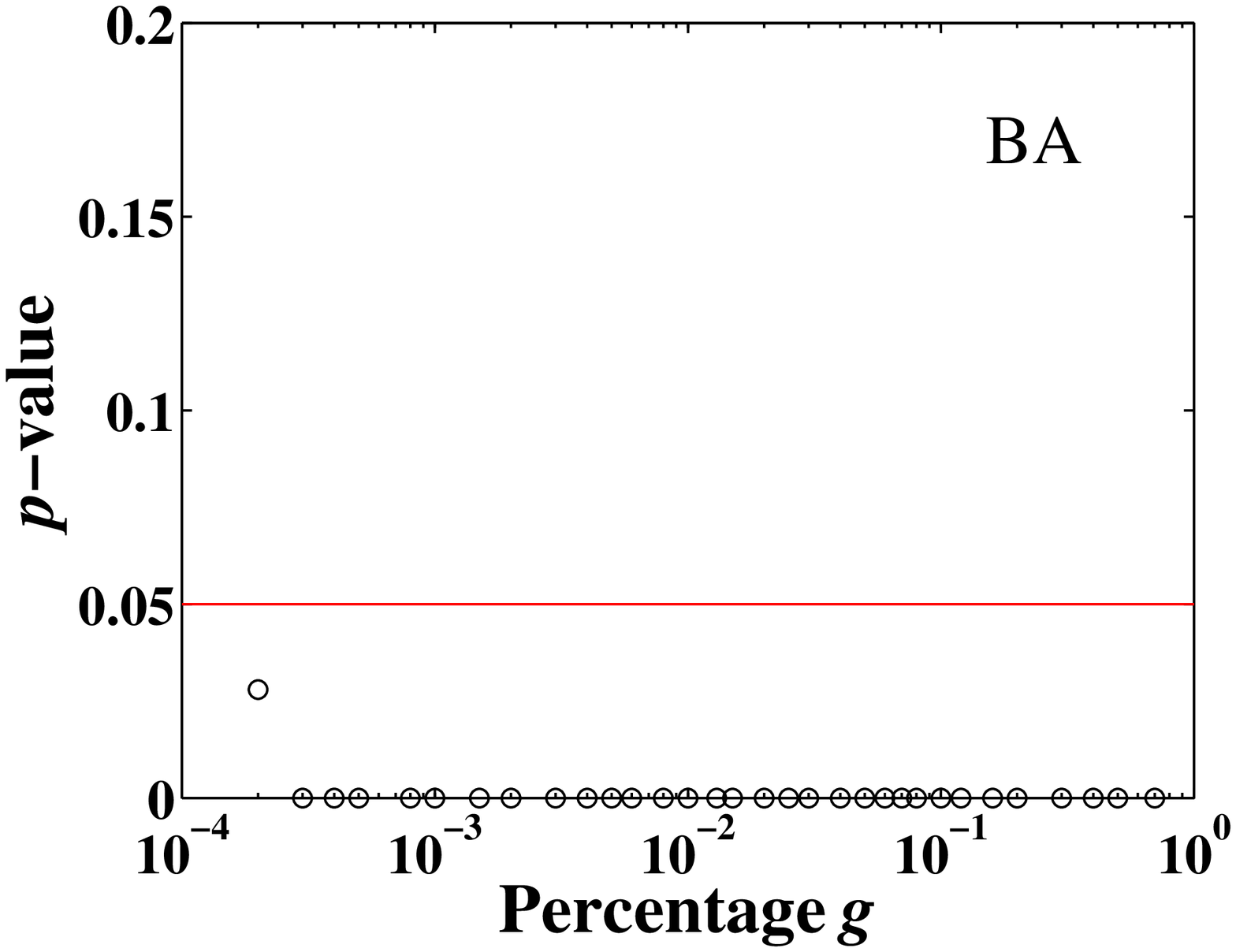}
\end{minipage}
\caption{Statistical tests for the presence of rich-club phenomena.
} \label{Fig:pvalue}
\end{figure*}

Figure \ref{Fig:pvalue} shows the $p$-values as a function of the
percentage $g$ of rich nodes for the networks investigated. For the
protein interaction network and the ER network, the $p$-values are
larger than $\alpha$ when $g<10\%$. Therefore, there is no rich-club
ordering in these two networks. For the Internet, except for the
point at the smallest $g$ and the point with $g=0$, all $p$-values
are well below $\alpha=5\%$, indicating significant rich-club
ordering in the Internet. For the scientific collaboration network,
the $p$-values are less than $\alpha=5\%$ for most values of $g$.
However, the most connected scientists corresponding to small $g$ do
not form a rich club. According to the top-right panel of
Fig.~\ref{Fig:rho}, the group of these most connected scientists has
relatively large normalized rich-club coefficient. What is the most
surprising is that the MR network and the BA network have
significant rich-club phenomena.
%
%

Among these cases, the presence of rich-club in the Internet has
stirred quite a few debates. In a recent work
\cite{Zhou-Mondragon-2007-XXX}, Zhou and Mondrag{\'o}n find that
there is a clique of rich nodes that are completely connected, which
is an undoubtable hallmark for the presence of rich club. We can put
further evidence for this argument. As illustrated in
Fig.~\ref{Fig:phi}, the rich-club coefficients $\phi$ are close to 1
when $g$ is small for the Internet, the MR model, and the BA model.
This means that the richest nodes in these networks are almost fully
connected. This validates the intuitive definition that a rich club
is a group of nodes with high degree that are intensely linked. A
statistical test puts further credit on the declaration of Zhou and
Mondrag{\'o}n for the presence of rich club in the Internet.

\begin{figure*}[htb]
\flushleft
\hspace{0.9cm}%
\includegraphics[width=4.5cm,height=4cm]{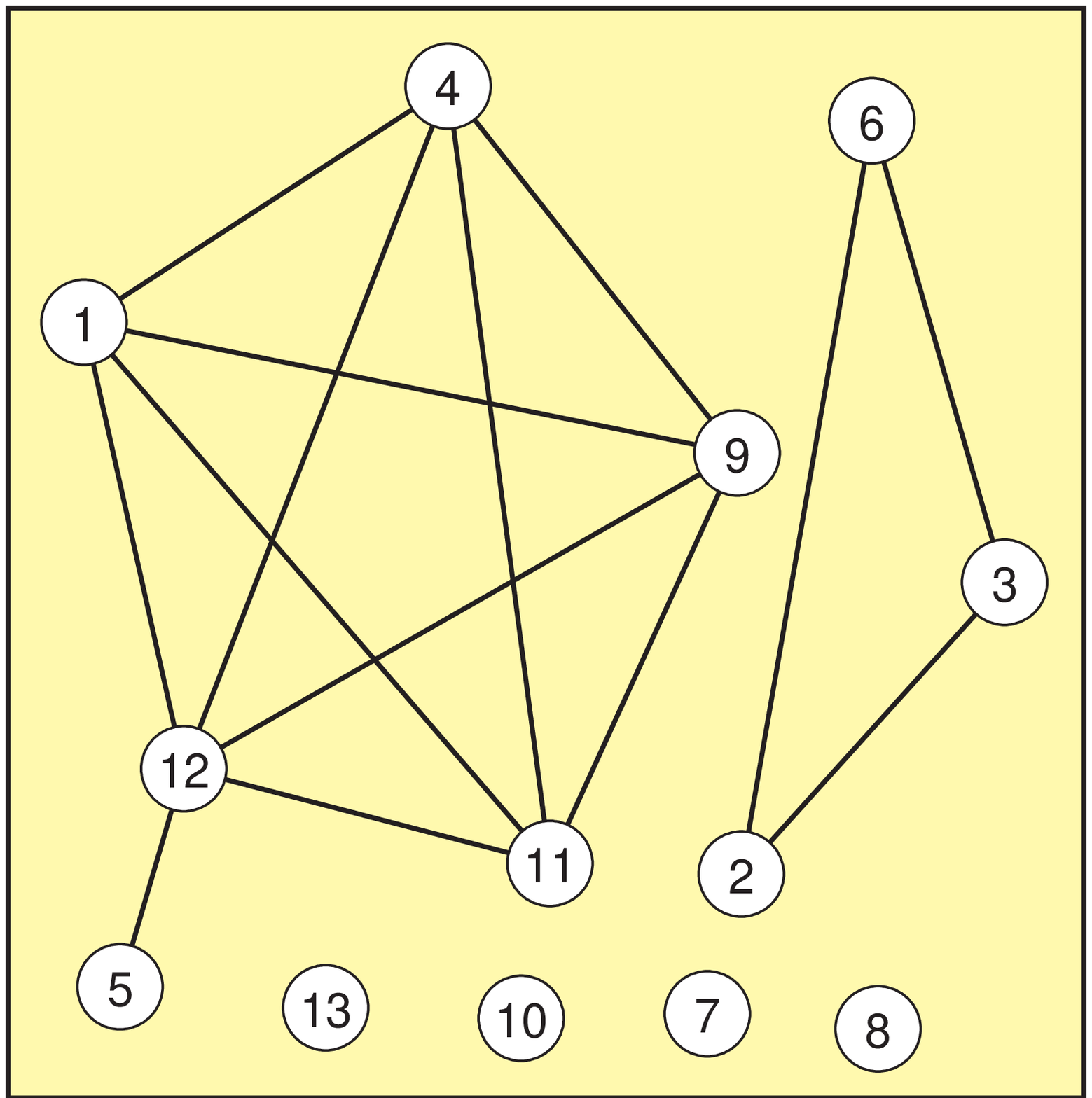}
\hspace{1.2cm}%
\includegraphics[width=4.5cm,height=4cm]{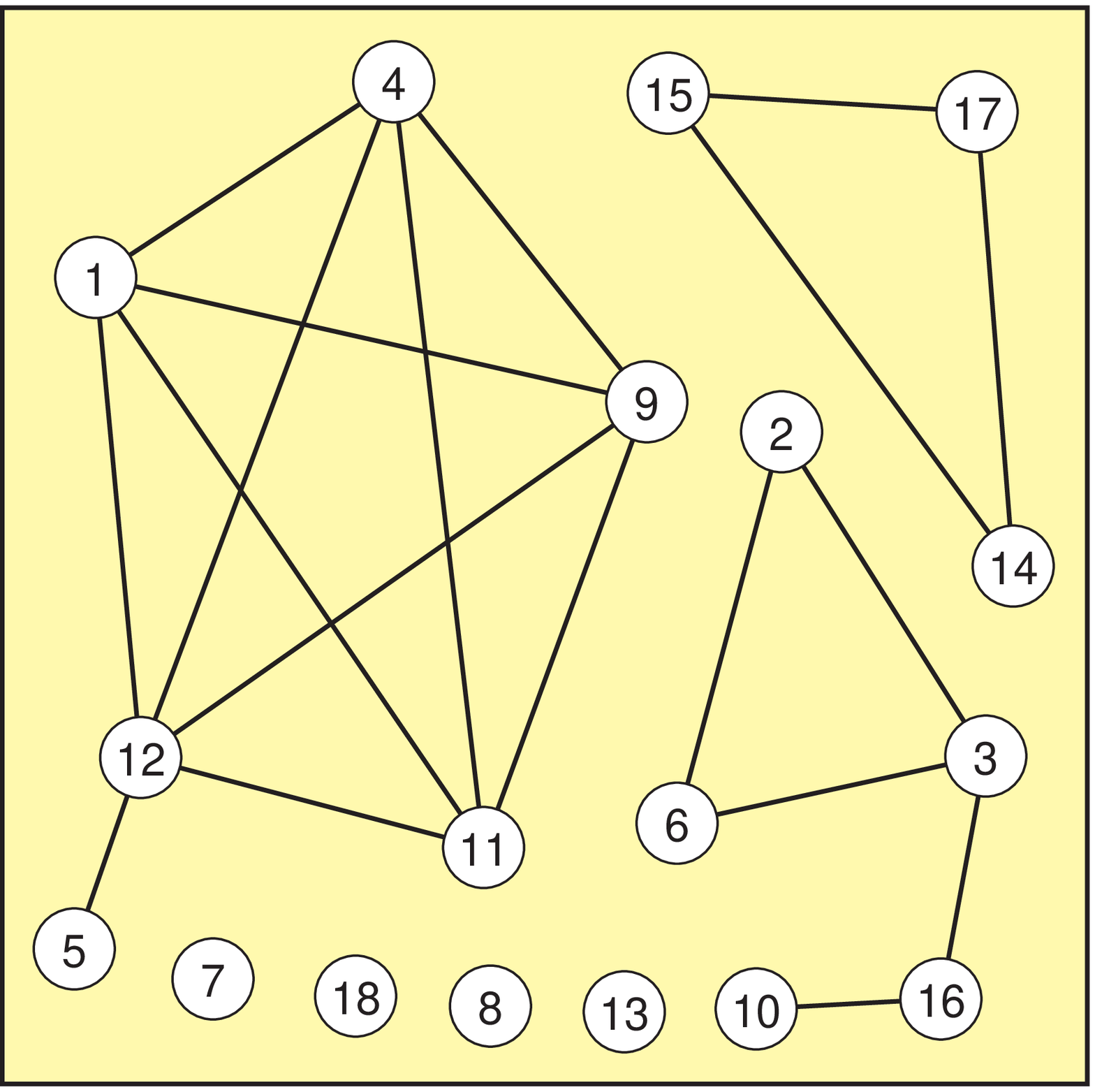}
\hspace{1.2cm}%
\includegraphics[width=4.5cm,height=4cm]{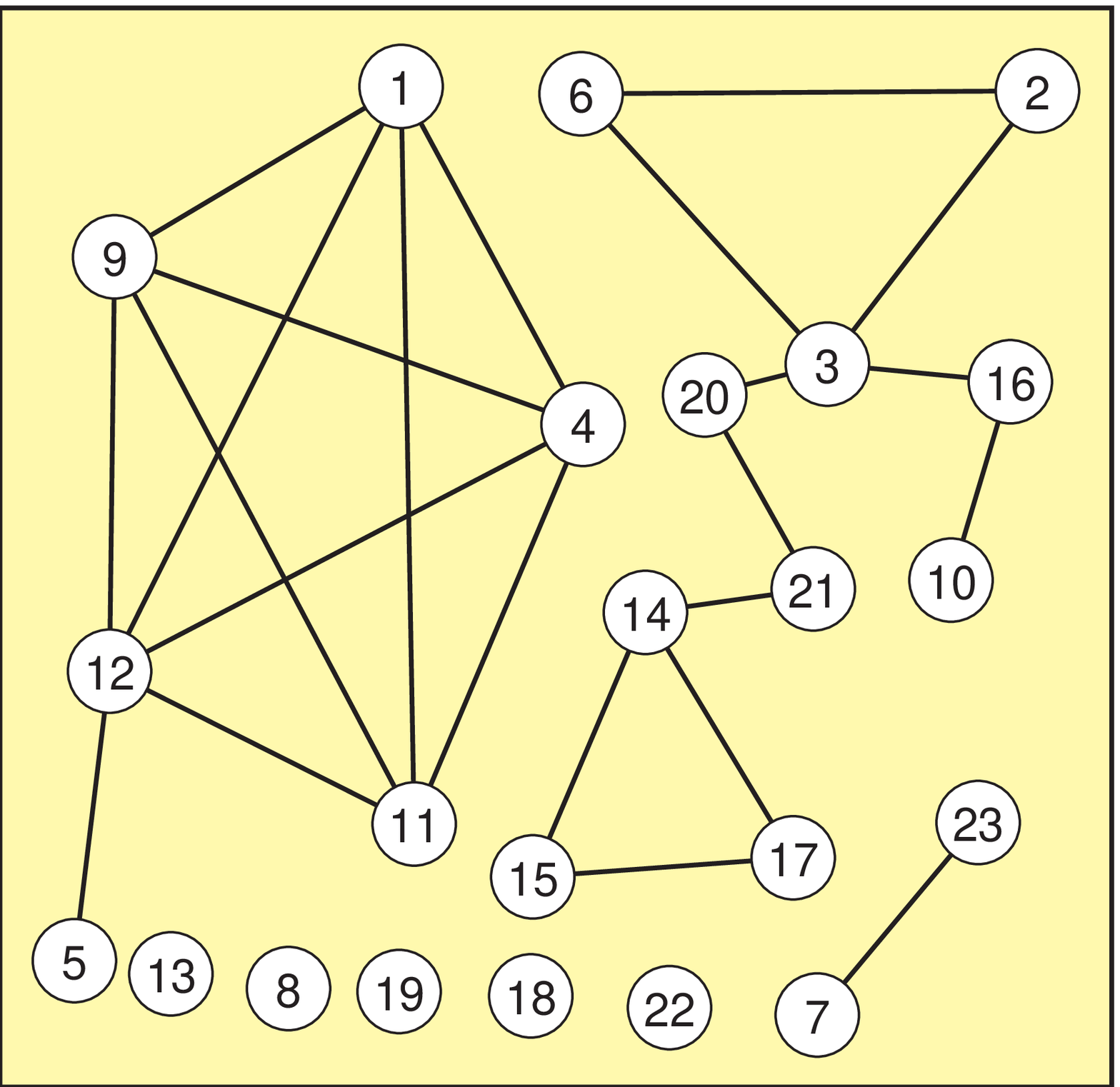}
\\
\vspace{0.2cm}
\begin{minipage}[t]{0.32\textwidth}
\includegraphics[width=5.4cm]{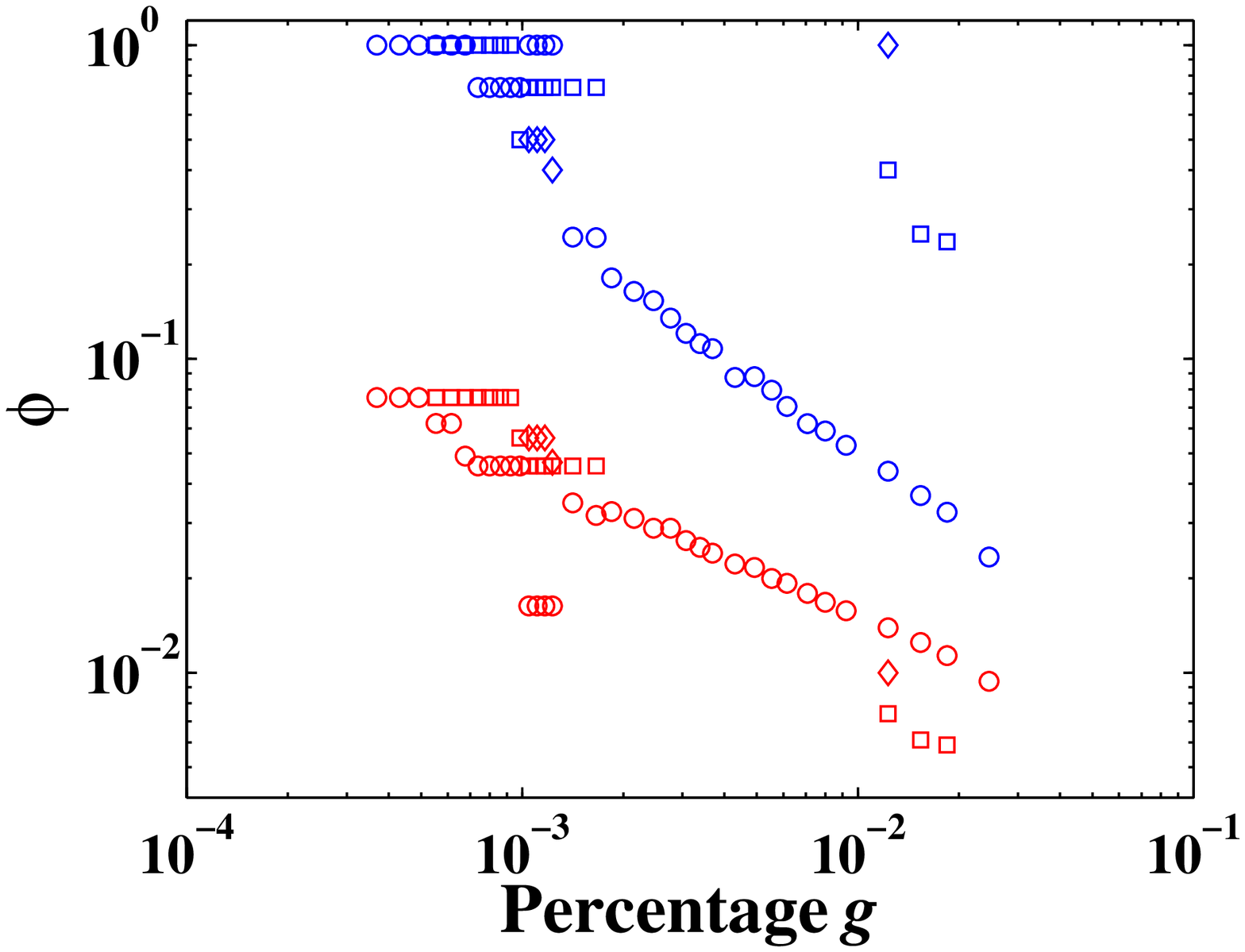}
\end{minipage}
\begin{minipage}[t]{0.32\textwidth}
\includegraphics[width=5.4cm]{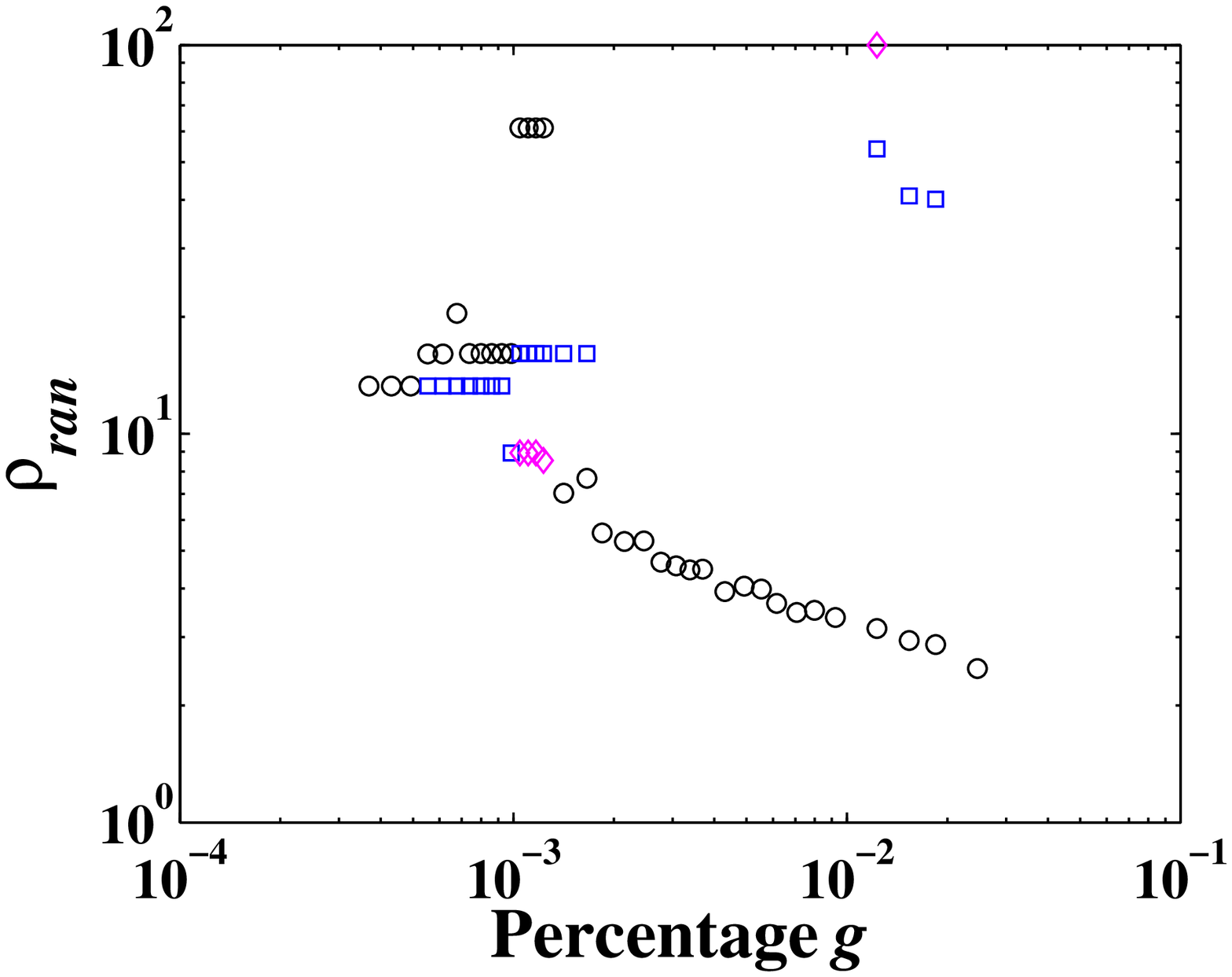}
\end{minipage}
\begin{minipage}[t]{0.32\textwidth}
\includegraphics[width=5.4cm]{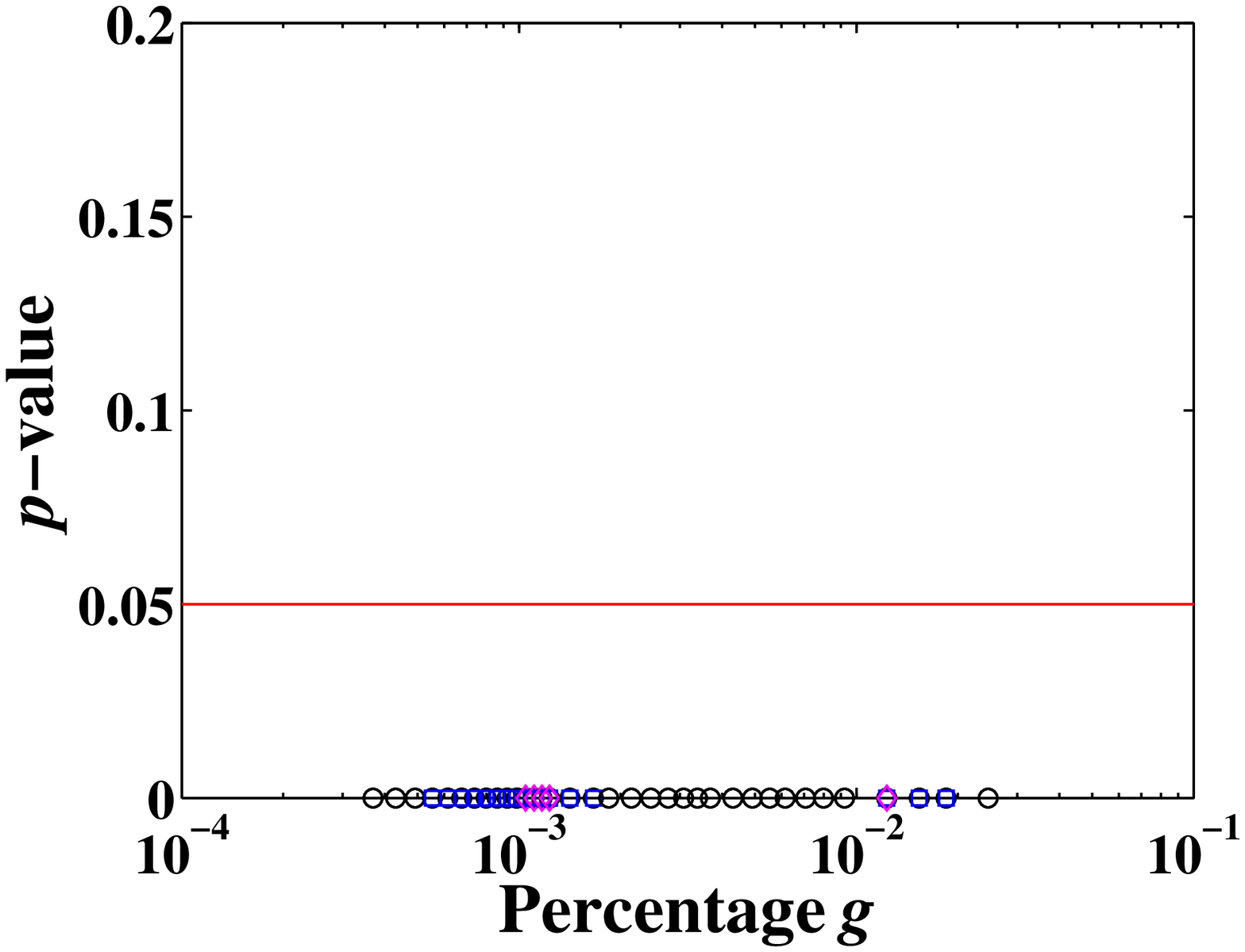}
\end{minipage}
\caption{(Color online) Reassessment of rich-club phenomena for the
scientific collaboration network. The upper panel shows the
subgraphs with $g=0.080\%$ ($k=66$), $g=0.111\%$ ($k=57$), and
$g=0.141\%$ ($k=54$). The lower panel shows the statistical analysis
on the sub-clubs for different $g$. The blue markers in the left
panel shows the rich-club coefficient $\phi$ for all isolated
sub-clubs with more than two nodes, while the red ones in the same
panel are the associated $\phi_{\rm{ran}}$. The middle panel
presents the $\rho$ function and the right panel digests the
corresponding $p$-values. It is observed that $p<\alpha$ for all
$g$.} \label{Fig:NEWpvalue}
\end{figure*}

A missing ingredient in the discussions of rich-club phenomenon is
the connectedness of the rich club. When we define rich nodes as
those with for example $g>1\%$ and start to investigate whether
these nodes form a rich club, a scrutiny should be carried out to
see if this ``club'' contains several disconnected sub-clubs. As
illustrated in the upper panel of Fig.~\ref{Fig:NEWpvalue}, the
scientific collaboration network are not fully connected for small
$g$. There are several separated clusters for small $g$. According
to Fig.~\ref{Fig:pvalue}, all these three subgraphs are rich clubs,
which however contradicts the common intuition that the members are
aware of each other forged by other members in the club. For
$g=0.141$, there are two rich clubs $(1,4,5,9,11,12)$ and
$(2,3,6,10,16,20,21,14,15,17)$. With the increase of richness
(smaller $g$ or larger $k$), the rich club $(1,4,5,9,11,12)$ remains
unchanged. The second rich club $(2,3,6,10,16,20,21,14,15,17)$
splits into two clubs $(2,3,6,10,16)$ and $(14,15,17)$ when node 20
and node 21 are removed for $g=0.111$. When $g=0.080$, the rich club
$(14,15,17)$ disappears and $(2,3,6,10,16)$ degenerates to
$(2,3,6)$. Therefore, when there are more than one isolated clusters
of nodes for a given $g$, we should investigate their statistical
significance one by one except for the trivial cases of isolated
nodes and pairs of nodes. The lower panel of
Fig.~\ref{Fig:NEWpvalue} shows the results for the scientific
collaboration network. One observes that $p<5\%$ for all clusters.

%
%

\begin{figure*}[htb]
\centering
\includegraphics[width=15cm]{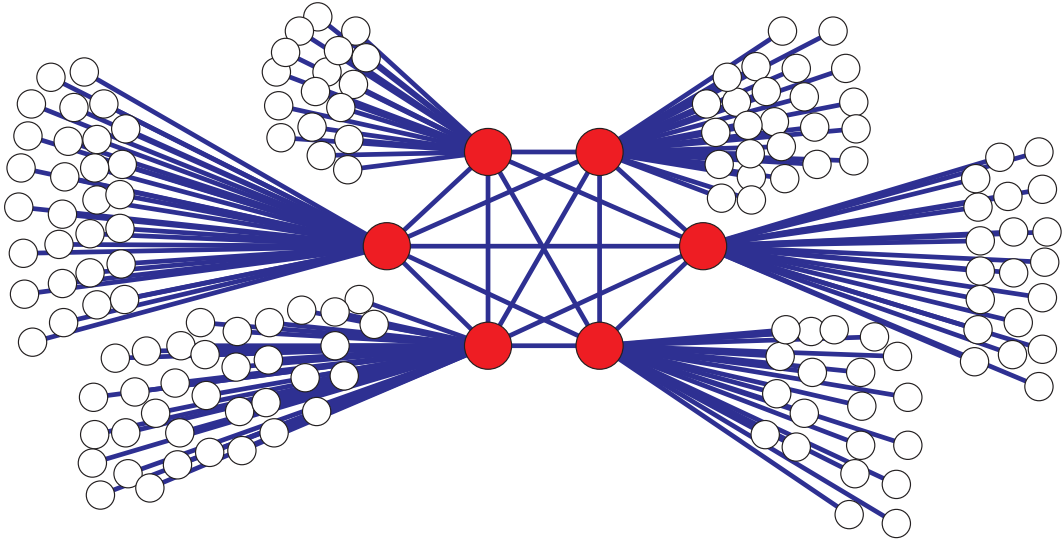}
\caption{(Color online) A schematic of illusionary rich-club
phenomenon. The big red nodes have larger degrees and form a
subnetwork which is completely connected. The rich-club coefficient
is $\phi(1)=1$. The network is disassortative with a Pearson
coefficient of $-0.77$. However, all its maximally random networks
have $\phi(1)=1$, which means that there is no rich-club phenomenon
statistically.} \label{Fig:FalseExample}
\end{figure*}

So far, we have shown that performing statistical test is necessary
which does a good job in the detection of rich clubs in complex
networks. However, a story always has two sides. Consider a toy
network shown in Fig.~\ref{Fig:FalseExample}. The graph consists of
two kinds of nodes identified with different colors: The degree of
each white node is $k=1$, while the red nodes are very ``rich'' and
fully connected. It is evident that the rich-club coefficient of the
red nodes is $\phi(k=1)=1$ and one would say they are within a rich
club without any doubt. Indeed, a qualitatively same figure was
taken as an example for the presence of rich club
\cite{Colizza-Flammini-Serrano-Vespignani-2006-NP}. Surprisingly,
this observation of $\phi(k=1)$ does not ensure that the red nodes
form a rich club in neither framework of $\rho>1$ adopted by Colizza
{\em{et al.}} \cite{Colizza-Flammini-Serrano-Vespignani-2006-NP} and
the statistical test proposed in this work since
$\phi_i(k=1)\equiv1$ for all maximally random networks. Hence, we
have $\rho(k=1)=1$, which means that there is no rich-club ordering
when $k=1$. This conclusion contradicts our intuitions.


We can generalize our discussion above by considering a network
consists of $m$ rich nodes, which are linked to $k_1$, $k_2$,
$\cdots$, $k_m$ nodes of degree $k=1$, respectively. Since each node
with $k=1$ has to be linked to a node with $k>1$ to ensure the
connectedness of the randomized network, the group of the $m$ rich
nodes have $\sum_{i=1}^mk_i$ out-edges and $E_{>1}$ edges among
them. The value of $E_{>1}$ does not change for all randomized
networks. In other words, $\phi_{\rm{ran}}(k=1)=\phi(k=1)$ and
$\rho(k=1)=1$. This class of artificial networks invalidates the
sophisticated approach based on statistical tests.

The analysis presented here provides a more rigorous methodology for
detecting rich clubs in complex networks. This allows us to
understand this phenomenon on a solid basis. However, there exist a
class of artificial networks with rich clubs on which the methods
based on null models taking maximally random networks. In this
sense, the definition of rich-club phenomenon remains an open
problem.

\begin{acknowledgments}
This work was partially supported by the National Natural Science
Foundation of China (Grant No. 70501011), the Fok Ying Tong
Education Foundation (Grant No. 101086), and the Shanghai
Rising-Star Program (Grant No. 06QA14015). We are grateful to thank
Dr. Newman for providing autonomous systems network data and
scientific collaboration network data.
\end{acknowledgments}

\bibliography{E:/Papers/Auxiliary/Bibliography} 

\newpage

\end{document}